\documentclass{article}

\usepackage[preprint]{neurips_2025}

\usepackage[utf8]{inputenc} 
\usepackage[T1]{fontenc}    
\usepackage{hyperref}       
\usepackage{url}            
\usepackage{booktabs}       
\usepackage{amsfonts}       
\usepackage{nicefrac}       
\usepackage{microtype}      
\usepackage{xcolor}         
\usepackage{amsmath}         

\usepackage{verbatim}
\usepackage{cuted}
\usepackage{mathtools, bm}
\usepackage{amssymb,amsfonts,latexsym,color}
\usepackage{algorithmic}
\usepackage{comment}
\usepackage{wrapfig}
\usepackage{makecell}

\usepackage[ruled,vlined,linesnumbered]{algorithm2e} %
\usepackage{float}
\usepackage{graphicx}
\usepackage{subcaption}  
\usepackage{caption}
\usepackage{amsmath}
\newtheorem{theorem}{Theorem}[section]
\newtheorem{lemma}[theorem]{Lemma}

\SetKwInput{KwInput}{Input}
\SetKwInput{KwOutput}{Output}

\title{Fair Cooperation in Mixed-Motive Games via Conflict-Aware Gradient Adjustment}

\author{%
  Woojun Kim \\
  Robotics Institute\\
  Carnegie Mellon University\\
  Pittsburgh, PA 15213 \\
  \texttt{woojunk@andrew.cmu.edu} \\
  \And
  Katia Sycara \\
  Robotics Institute\\
  Carnegie Mellon University\\
  Pittsburgh, PA 15213 \\
  \texttt{sycara@andrew.cmu.edu} \\
}

\begin{document}

\maketitle

\begin{abstract}
  Multi-agent reinforcement learning in mixed-motive settings presents a fundamental challenge: agents must balance individual interests with collective goals, which are neither fully aligned nor strictly opposed. To address this, reward restructuring methods such as gifting and intrinsic motivation have been proposed. However, these approaches primarily focus on promoting cooperation by managing the trade-off between individual and collective returns, without explicitly addressing fairness with respect to agents’ task-specific rewards. In this paper, we propose an adaptive conflict-aware gradient adjustment method that promotes cooperation while ensuring fairness in individual rewards. The proposed method dynamically balances policy gradients derived from individual and collective objectives in situations where the two objectives are in conflict. By explicitly resolving such conflicts, our method improves collective performance while preserving fairness across agents. 
  We provide theoretical results that guarantee monotonic non-decreasing improvement in both the collective and individual objectives and ensure fairness. Empirical results in sequential social dilemma environments demonstrate that our approach outperforms baselines in terms of social welfare, while maintaining fairness.
\end{abstract}

\section{Introduction}

Multi-agent reinforcement learning (MARL) aims to train multiple agents to maximize cumulative rewards in a given task. Depending on the reward structure, MARL is typically categorized into three settings: cooperative, adversarial, and mixed-motive. In the mixed-motive setting, agents' rewards are neither fully aligned (as in cooperative settings) nor entirely opposed (as in adversarial settings), necessitating that each agent balances self-interest with the collective interest. This mixed-motive setting is frequently encountered in real-world applications. For example, in traffic control systems, each agent (e.g., a local intersection controller) may aim to minimize local congestion, which can conflict with global traffic flow optimization if not coordinated. A similar tension happens in sequential social dilemmas (SSDs) such as Cleanup or Harvest~\cite{leibo2017multi}, where agents must invest in public goods (e.g., cleaning waste or harvesting resources judiciously) that benefit the group but do not yield immediate individual rewards.

However, achieving such a balance in mixed-motive settings is inherently challenging. Excessively selfish behavior by agents can deteriorate collective welfare, which, in turn, negatively impacts each agent’s own return---creating a vicious cycle that ultimately harms all participants. Additionally, in some scenarios, certain agents must sacrifice their own returns to improve the collective outcome, potentially leading to unfairness. Conversely, an excessive focus on fairness can hinder learning in tasks that require cooperation. Therefore, it is crucial to enhance collective outcome while ensuring fairness by appropriately balancing individual and collective interests.

In mixed-motive settings, many approaches adopt reward restructuring by incorporating intrinsic rewards such as social influence~\cite{jaques2019social}, formal contracts~\cite{haupt2024formal}, gifting~\cite{lupu2020gifting, konglearning}, and inequity aversion~\cite{hughes2018inequity}. These methods primarily aim to maximize the collective return by mediating the trade-off between self-interest and collective interests. For example, gifting mechanisms promote cooperation by enabling agents to share a portion of their rewards with others. However, despite their effectiveness in inducing cooperation, such reward restructuring may raise fairness concerns, for example, the gifted reward is intrinsic and not part of the task-defined reward that agents are fundamentally trained to maximize. Consider the Cleanup environment: agents only receive extrinsic rewards for collecting apples, yet apples will only regrow if waste is cleaned. It is often observed that some agents specialize in cleaning waste while others collect apples and subsequently gift a portion of their reward to those who sacrificed their own gain. Although this leads to improved collective performance, the agents engaged in waste cleaning never directly receive task rewards from apple collection. This becomes even worse if the agents are trained with the collective return, since some agents are encouraged to clean the waste all the time. Aside from reward restructuring, an approach has been proposed to align individual and collective objectives by adjusting policy gradients toward stable fixed points of the collective return, while still considering individual interests~\cite{lialigning}. However, this method does not adequately consider fairness, as it primarily focuses on stability without explicitly addressing the conflict between individual and collective objectives.

In order to enhance cooperation while ensuring fairness, we propose a fair and conflict-aware gradient adjustment method (FCGrad) that dynamically balances gradients derived from individual and collective objectives by explicitly handling conflicts between them. FCGrad first detects the presence of conflicts, and when conflicts are found, it projects one gradient onto the normal plane of the other---preserving one objective’s direction while avoiding interference with the other. Notably, FCGrad prioritizes the gradient associated with the lower objective value. For example, if the individual objective is lower than the collective objective, indicating that the agent is in an unfair situation, we project the individual gradient onto the normal plane of the collective gradient and use the result as the final update. This enables cooperation to be enhanced while maintaining fairness by resolving conflicts. We provide theoretical results showing that, under certain assumptions, the proposed gradient method guarantees monotonic non-decreasing improvement in both collective and individual objectives. We further show that the two objectives converge to the same value, leading to all agents' objectives aligning---thus ensuring individual fairness. In addition, we empirically demonstrate the effectiveness of FCGrad in terms of $\alpha$-fairness~\cite{mo2000fair}, which captures both performance and fairness, in the Unfair Coin Game and two sequential social dilemma environments: Cleanup and Harvest.

\section{Background and Related Works}

\subsection{Partially Observable Stochastic Game}

A \textit{Partially Observable Markov Game} (POMG) models multi-agent decision-making under uncertainty~\cite{littman1994markov, emery2004approximate}. A POMG is defined as a tuple \( (N, S, \{A_i\}_{i=1}^N, T, \{O_i\}_{i=1}^N, \{R_i\}_{i=1}^N, \gamma) \), where \(N\) is the number of agents, \(S\) is the set of states, \(A_i\) is the action set of agent \(i\), \(T: S \times A_1 \times \cdots \times A_N \rightarrow \Delta(S)\) is the transition function, \(O_i: S \rightarrow \Delta(\mathcal{O}_i)\) is the observation function, \(R_i: S \times A_1 \times \cdots \times A_N \rightarrow \mathbb{R}\) is the reward function for agent \(i\), and \(\gamma \in [0,1)\) is the discount factor. Here, depending on the reward structure, a POMG can represent various types of multi-agent settings: \textit{cooperative} settings~\cite{jeon2022maser,  kim2023variational, kim2023adaptive, kim2023parameter}, where all agents share an identical reward function (i.e., \(r^1 = \cdots = r^N\)); \textit{adversarial} settings~\cite{hansolution, yu2019multi}, where agents have directly opposing objectives, often modeled as zero-sum (i.e., \(\sum_{i=1}^N r^i = 0\)); or \textit{mixed-motive} settings~\cite{mckee2020social, konglearning}, where agents' rewards are neither fully aligned nor strictly opposed, creating simultaneous incentives for both cooperation and competition.


\subsection{Mixed-motive Coordination in Multi-Agent RL}

We consider mixed-motive settings, where agents' self-interest often conflicts with collective outcomes. Let us define the \textit{collective return} as the average of \textit{individual returns}: $R_{col}=\frac{1}{N} \sum_{i=1}^N R^i(s, a)$, where $R^i(s,a)$ is the individual return of Agent $i$. In the context of gradient-based learning, a conflict occurs when the local and collective return gradients are misaligned, that is, when $\nabla_{\theta_i} \mathbb{E}\left[R^i\right] \cdot \nabla_{\theta_i} \mathbb{E}\left[R_{\text{col}}\right] < 0,$ where $\theta_i$ denotes the parameters of Agent $i$'s policy. 

To enhance cooperation (i.e. maximize collective reward) while avoiding conflicts, a variety of approaches have been proposed, including inequity aversion~\cite{hughes2018inequity, wang2019evolving}, social influence~\cite{jaques2019social}, reciprocal reward shaping~\cite{zhou2024reciprocal}, formal contract mechanisms~\cite{haupt2024formal}, and gifting-based cooperation~\cite{lupu2020gifting, konglearning}. Many of these approaches are studied in the context of \textit{Sequential Social Dilemmas (SSDs)}~\cite{leibo2017multi}, a prominent class of mixed-motive settings in which agents repeatedly arbitrate between short-term selfish actions and long-term collective returns. For example, \cite{konglearning} proposed a gift-based method that balances altrusim and self-interest based based on social relationships between agents. \cite{jaques2019social} proposed an intrinsic motivation method that rewards agents for exerting causal influence over others’ actions, thereby improving coordination in SSDs. \cite{hughes2018inequity} introduced inequity-averse agents that learn to cooperate by assigning temporal credit to prosocial behavior and penalizing inequitable outcomes. The aforementioned methods can be broadly viewed as forms of reward shaping, wherein additional intrinsic or socially-informed rewards guide agents toward cooperative behavior.

In contrast to reward shaping approaches, recent work~\cite{lialigning} has explored direct optimization in the gradient space to reconcile individual and collective objectives. Specifically, the Altruistic Gradient Adjustment (AgA) method~\cite{lialigning} modifies the policy gradients of both the collective and individual objectives, pulling agents toward stable fixed points of the collective objective and pushing them away from unstable ones. The adjusted gradient for Agent $i$ is defined as $g^i_{aga} = g_{col} + \lambda (g^i_{ind} +H_{col}^T g_{col})$, where $g_{col}$ and $g_{ind}$ are the gradients of the collective and individual objectives for Agent $i$, $H_{col}^T$ is the Hessian of the collective return with respect to the policy parameters, and $\lambda$ is the adjustment coefficient and its sign is determined by $\text{sign}[(g_{\mathrm{col}} \cdot H_{col}^T  g_{col})\left[ (g^i_{ind}\cdot H_{col}^T g_{col})+\|H_{col}^T \cdot g_{col}\|^2 \right]]$. This adjustment steers the update direction according to the local stability of the collective objective.
Despite its effectiveness, AgA incurs additional computational complexity, focuses on the stability of the collective objective rather than directly resolving gradient conflicts, and provides no guarantees of monotonic improvement or fairness.

\subsection{Gradient Adjustment}

Gradient adjustment approaches have been actively investigated in multi‑task learning~\cite{yu2020gradient, liu2021conflict, navon2022multi, senushkin2023independent}. For example, CAGrad~\cite{liu2021conflict} formulates a quadratic program to compute a conflict‑averse convex combination of gradients, achieving better trade‑offs at the cost of increased complexity, and Nash‑MTL~\cite{navon2022multi} frames the task‑weighting problem as a bargaining game, using the Nash bargaining solution to promote fairness and efficiency across tasks. Another method that inspires our work is PCGrad~\cite{yu2020gradient}, which mitigates conflicts by projecting each conflicting gradient onto the normal plane of the other, offering a simple yet effective solution with low computational overhead. Specifically, when two gradients $g_1$ and $g_2$ are conflicted, PCGrad adjusts them by projecting one onto the normal plane of the other, i.e., $\tilde{g}^{PCGrad}_1=g_1-\frac{g_1\cdot g_2}{\|g_2\|^2}g_2$, and then uses the average of $\tilde{g}^{PCGrad}_1$ and $\tilde{g}^{PCGrad}_2$ as the final update.


\subsection{Fairness in Multi-agent RL}

Fairness concerns how returns are distributed among agents rather than how large the total return is, making it complementary, but often orthogonal to cooperation and efficiency. Fairness has been considered in multi-agent RL literature in both cooperative and mixed-motive settings~\cite{zimmer2021learning, grupen2022cooperative, aloor2024cooperation, smit2024learning}. For example, in cooperative settings, \cite{zimmer2021learning} formulates fairness as the optimization of a fair social welfare function and \cite{grupen2022cooperative} proposes a method for achieving team fairness by enforcing permutation-equivariant policies, which mitigate emergent unfairness caused by asymmetric role assignment. In mixed‑motive settings, 
\cite{konglearning} shows enhanced fairness when measuring the sum of individual rewards and gifts, whereas in this paper we evaluate fairness using individual rewards only. \cite{hughes2018inequity}, inspired by the literature on inequality in economics~\cite{engelmann2004inequality}, explicitly leverages fairness by adding both disadvantage and advantage inequality terms to the reward of each agent to improve cooperation in SSD. Specifically, the shaped reward for Agent is $r^i=r^i-\alpha_{IA}/(N-1)\sum_{j\neq i}\text{max}(r_j-r_i, 0)-\beta_{IA}/(N-1)\sum_{j\neq i}max(r_i-r_j, 0)$, where $\alpha_{IA}$ and $\beta_{IA}$ weight disadvantage and advantage inequity, respectively.

\section{Methodology}

In mixed-motive settings, the individual and collective objectives may be either aligned or in conflict. When they are aligned, optimizing both objectives is sufficient, as neither interferes with the other. In such cases, an appropriately weighted combination of the two can be effective. However, when the objectives are in conflict, it becomes essential to explicitly address the interference between them, as prioritizing one may hinder the other. This is because focusing solely on the individual objective may hinder learning in tasks where cooperative behavior is essential for maximizing individual returns, while focusing solely on the collective objective may compromise fairness among agents. Therefore, it is important to (1) recognize when such conflicts arise and (2) correspondingly adjust the individual and collective objectives, appropriately considering both fairness and cooperation.

\begin{figure}[t]
\centering

\begin{minipage}[t]{0.48\textwidth}
\centering
\vspace{-15ex}
\begin{algorithm}[H]
\caption{FCGrad}
\label{alg:fcgrad}
\KwIn{Policy parameters $\theta$, learning rate $\eta$, weighting factor $\beta$}

Compute $g_{\text{ind}} := \nabla_\theta V_{\text{ind}}(\theta)$, $g_{\text{col}} := \nabla_\theta V_{\text{col}}(\theta)$\

\eIf{$\langle g_{\text{ind}}, g_{\text{col}} \rangle \geq 0$}{
    $g_{\smash{\raisebox{-0.7ex}{\scriptsize\text{FCGrad}}}} \gets (1 - \beta) g_{\text{ind}} + \beta g_{\text{col}}$\;
}{
    \eIf{$V_{\text{col}}(\theta) \geq V_{\text{ind}}(\theta)$}{
        $g_{\smash{\raisebox{-0.7ex}{\scriptsize\text{FCGrad}}}} \gets g_{\text{ind}} - \dfrac{\langle g_{\text{col}}, g_{\text{ind}} \rangle}{\|g_{\text{col}}\|^2} g_{\text{col}}$\;
    }{
        $g_{\smash{\raisebox{-0.7ex}{\scriptsize\text{FCGrad}}}} \gets g_{\text{col}} - \dfrac{\langle g_{\text{ind}}, g_{\text{col}} \rangle}{\|g_{\text{ind}}\|^2} g_{\text{ind}}$\;
    }
}
\textbf{Return} $\theta \gets \theta + \eta g_{\smash{\raisebox{-0.7ex}{\scriptsize\text{FCGrad}}}}$\;
\end{algorithm}
\end{minipage}%
\hfill
\begin{minipage}[t]{0.48\textwidth}
\centering
\begin{minipage}[t]{0.55\textwidth}
    \includegraphics[width=\linewidth]{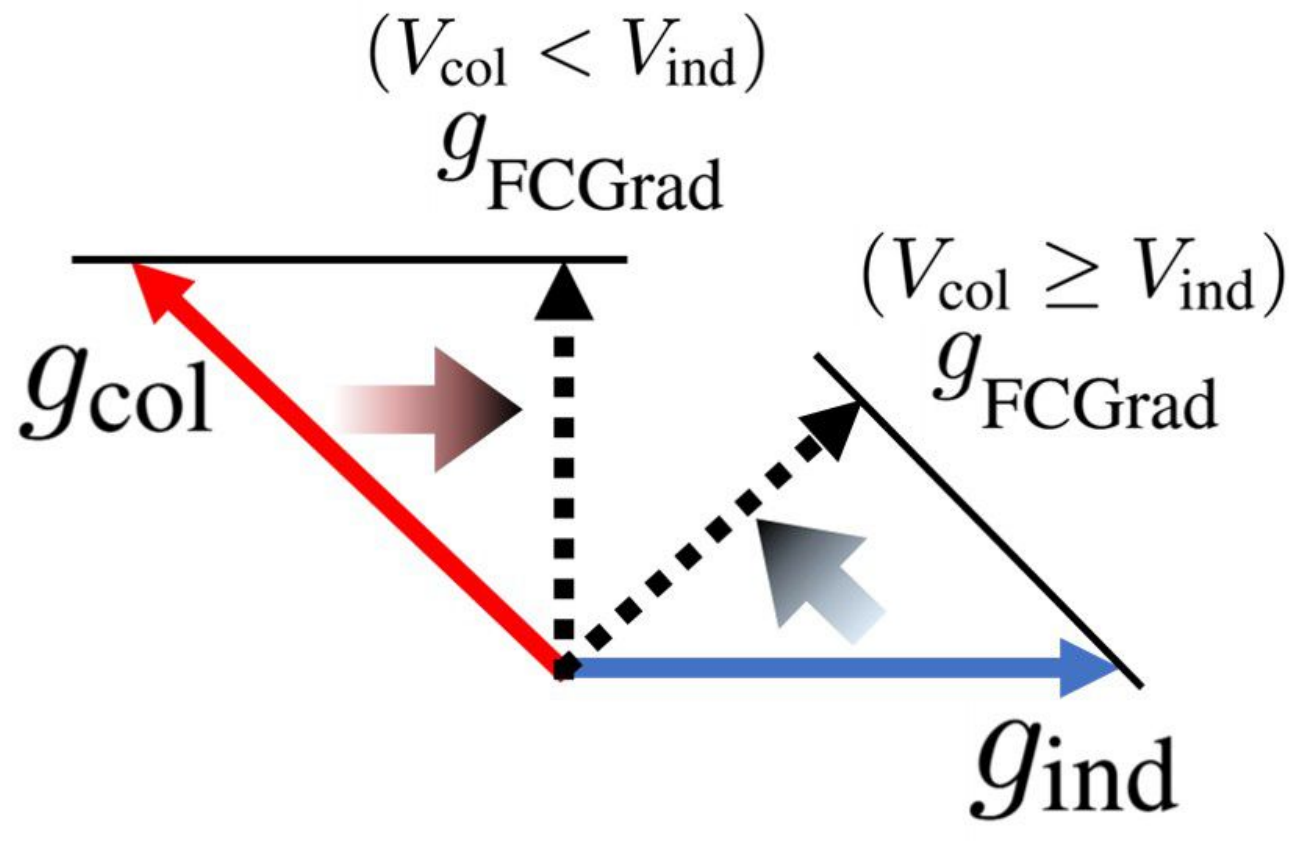}
    \caption*{(a) Conflict}
\end{minipage}%
\hfill
\begin{minipage}[t]{0.33\textwidth}
    \hspace{-0ex}\includegraphics[width=\linewidth]{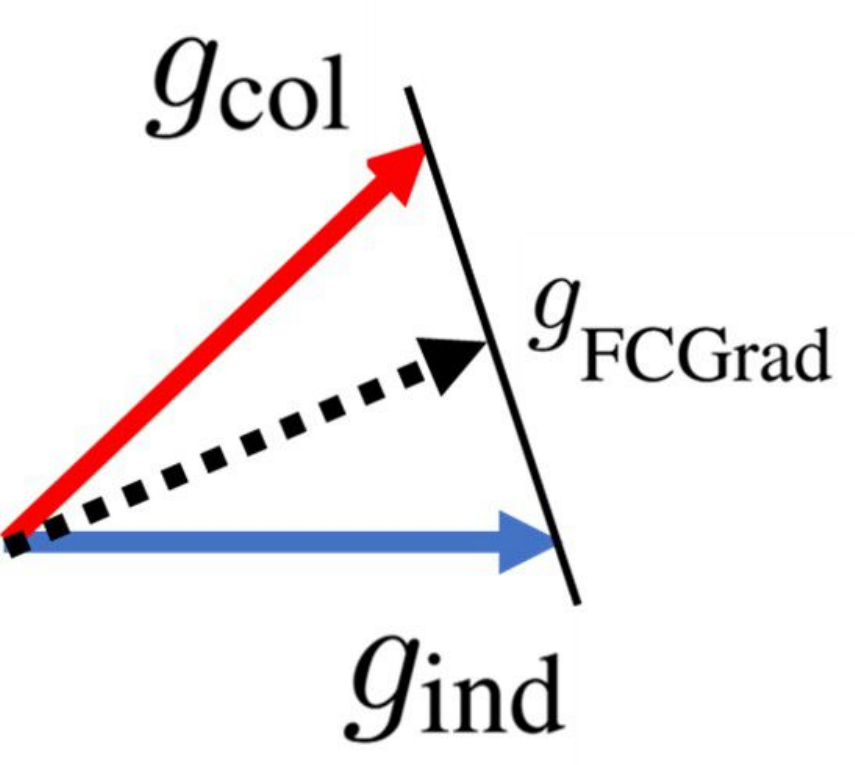}
    \caption*{\hspace{-2ex}(b) Non-conflict}
\end{minipage}
\vspace{1ex}
\captionof{figure}{FCGrad illustration: (a) When conflicts occur, the gradient corresponding to the lower objective---either individual or collective---is projected onto the normal plane of the gradient of the higher objective; (b) When no conflict is detected, a task-dependent weighted sum of the two gradients is applied.
}
\label{fig:fcgrad_full}
\end{minipage}

\end{figure}

To this end, we propose a fair and conflict-aware gradient adjustment method, called FCGrad, which guarantees the monotonic non-decrease of both individual and collective objectives, while preserving fairness across individual objectives. Specifically, when the individual and collective gradients are in conflict, FCGrad projects the gradient associated with the lower expected return onto the normal plane of the other. This projected gradient remains a valid ascent direction for its own objective while avoiding interference with the other, and is then used as the final update. For example, if the individual expected return is lower than the collective expected return, indicating that the agent is disadvantaged in terms of fairness, we project the gradient of the individual objective onto the normal plane of the collective gradient and use it as the update direction. The detailed procedure and a visual illustration of FCGrad are provided in Algorithm~\ref{alg:fcgrad} and Fig.~\ref{fig:fcgrad_full}, respectively.
In the following, we present the detailed method along with its theoretical analysis.



\subsection{FCGrad: Fair and Conflict-aware Gradient Adjustment}

We now describe how FCGrad operates from the perspective of Agent~$i$. Let $\theta \in \mathbb{R}^d$ denote the parameters of the policy $\pi_\theta$ for Agent~$i$. Let us define $V_{\text{ind}}(\theta)$ and $V_{\text{col}}(\theta)$ as the expected individual and collective returns, respectively, computed under the initial state distribution. Note that $V_{\text{ind}}(\theta)$ and $V_{\text{col}}(\theta)$ are the individual and collective objectives, respectively. Let $g_{\text{ind}} := \nabla_\theta V_{\text{ind}}(\theta)$ and $g_{\text{col}} := \nabla_\theta V_{\text{col}}(\theta)$ denote the gradients of the individual and collective objectives, respectively. These represent ascent directions for $V_{\text{ind}}(\theta)$ and $V_{\text{col}}(\theta)$, meaning that for a sufficiently small $\eta > 0$, the following holds: $V_{\text{ind}}(\theta + \eta g_{\text{ind}}) > V_{\text{ind}}(\theta)$ and $V_{\text{col}}(\theta + \eta g_{\text{col}}) > V_{\text{col}}(\theta)$.

FCGrad proceeds as follows:  
(1) check whether $g_{\text{ind}}$ and $g_{\text{col}}$ are in conflict by examining the sign of their inner product, where a negative inner product indicates the presence of a conflict. (2) if $\langle g_{\text{ind}}, g_{\text{col}} \rangle \geq 0$ (i.e., non-conflict), FCGrad uses the weighted sum of two gradients: $g = (1-\beta) g_{\text{ind}} +\beta g_{\text{col}}$, (3) $\langle g_{\text{ind}}, g_{\text{col}} \rangle < 0$ (i.e., conflict), FCGrad places more weight on the individual (collective) gradient when the collective (individual) objective is greater, in order to ensure fairness. The corresponding gradient is given by
\begin{align}\label{eq:ourgradient}
g_{\smash{\raisebox{-0.7ex}{\scriptsize\text{FCGrad}}}}=
\begin{cases}
\tilde{g}_{ind} & \text{if  } (V_{\text{col}} \geq V_{\text{ind}}) \\
\tilde{g}_{col} & \text{if  } (V_{\text{col}} < V_{\text{ind}})
\end{cases}
\end{align}
where $\tilde{g}_{\text{col}}$ and $\tilde{g}_{\text{ind}}$ are the projections of $g_{\text{col}}$ and $g_{\text{ind}}$, respectively, onto the normal plane of another gradient vector, given by
\begin{align}
    \tilde{g}_{\text{col}} : =g_{\text{col}} - \frac{\langle g_{\text{ind}}, g_{\text{col}}\rangle}{\|g_{\text{ind}}\|^2}g_{\text{ind}}, ~~~~~~~~~~~ \tilde{g}_{\text{ind}} : =g_{\text{ind}} - \frac{ \langle g_{\text{col}} , g_{\text{ind}}\rangle}{\|g_{\text{col}}\|^2}g_{\text{col}}
\end{align}

(4) update the policy parameter with the step size $\eta$: $\theta \leftarrow \theta + \eta g$. Note that $\tilde{g}_{\text{col}}$ projects $g_{\text{col}}$ onto the normal plane of $g_{\text{ind}}$. Thus, $\tilde{g}_{\text{col}}$ is still a valid ascent direction for the collective objective while preserving the individual reward. This indicates that FCGrad prioritizes the individual objective without compromising the collective one when the agent is in an unfair situation, i.e., when the individual objective is lower. Conversely, when the collective objective is lower, FCGrad prioritizes the collective objective without compromising the individual one.

\subsection{Theoretical Analysis}

In this section, we prove that FCGrad guarantees monotonically non-decreasing improvements in both the collective and individual objectives, and that both objectives converge to the same value. This ensures that the expected individual returns across agents also converge to the same value.

\begin{theorem}
\label{theorem:increase}
Assume $V_{\text{ind}}(\theta)$ and $V_{\text{col}}(\theta)$ are differentiable and L-smooth. Let the update direction $g$ be defined as in Equation~\ref{eq:ourgradient}. Then, for a sufficiently small step size $\eta>0$, the update $\theta \leftarrow \theta + \eta g$ yields monotonically non-decreasing improvements in both $V_{\text{col}}(\theta)$ and $V_{\text{int}}(\theta)$.
\end{theorem}

\textit{Proof}. See Appendix A.

Theorem~\ref{theorem:increase} states that FCGrad ensures the monotonic non-decreasing in both $V_{\text{ind}}(\theta)$ and $V_{\text{col}}(\theta)$, under certain assumptions. Note that all agents are updated using FCGrad, so both the individual objectives of all agents and the collective objective, defined as the expected return averaged across agents, are improved accordingly.




\begin{theorem}
\label{theorem:gap-conv}
Assume $V_{\text{ind}}(\theta)$ and $V_{\text{col}}(\theta)$ be $L$-smooth, and let $\delta_t := V_{\text{ind}}(\theta_t) - V_{\text{col}}(\theta_t)$ denote the value gap at iteration $t$.  
Assume the step size satisfies the Robbins–Monro conditions:  
$0<\eta_t \leq |\delta_t|/L$ with $\sum_t \eta_t = \infty$ and $\sum_t \eta_t^2 < \infty$.  
Also assume conflict recurrence, meaning that for any $\epsilon > 0$ and any $t$, if $|\delta_t| \geq \epsilon$, then there exists $t' \geq t$ such that $(g_{\text{ind}, t'} \cdot g_{\text{col}, t'}) < 0$.  
Then, the value gap converges to zero:
\begin{align}
    \lim_{t \to \infty} |V_{\text{ind}}(\theta_t) - V_{\text{col}}(\theta_t)| = 0.
\end{align}
\end{theorem}

\textit{Proof.} See Appendix A.

Theorem~\ref{theorem:gap-conv} states that the gap between the collective and individual objectives converges to zero under certain assumptions, including conflict recurrence, where conflicts occur continuously. This assumption is reasonable in mixed-motive settings, especially near equilibrium, because agents face inherent tensions between cooperation and self-interest, and as they approach equilibrium, misalignments in their objectives can continue to induce conflicts, even with small policy updates. Under the assumption that all agents use FCGrad, the individual objectives of all agents converge to the collective objective, and thus all individual objectives converge to the same value. This, in turn, implies that individual fairness is achieved.


\subsection{Practical Algorithm}

We now introduce a practical FCGrad-based multi-agent RL algorithm for mixed-motive settings. We consider decentralized training and execution, where each agent does not have access to other agents' information but shares rewards, as commonly assumed in gifting mechanisms~\cite{lupu2020gifting, konglearning}. Each agent trains its policy and value functions for both individual and collective returns solely based on its own local observations and the shared rewards. For this, we construct two separate value networks for the individual and collective objectives, while sharing a common encoder between them. Each value function is trained using generalized advantage estimation~\cite{schulman2015high} to compute the corresponding advantage estimates. Using the two value functions, we compute the policy gradients of the individual and collective objectives, denoted as $g_{\text{ind}}$ and $g_{\text{col}}$, respectively, via the PPO policy gradient. These gradients are then combined using FCGrad to determine the final update direction.


\begin{figure}[t!]
\begin{tabular}{ccc}
\includegraphics[width=0.30\linewidth]{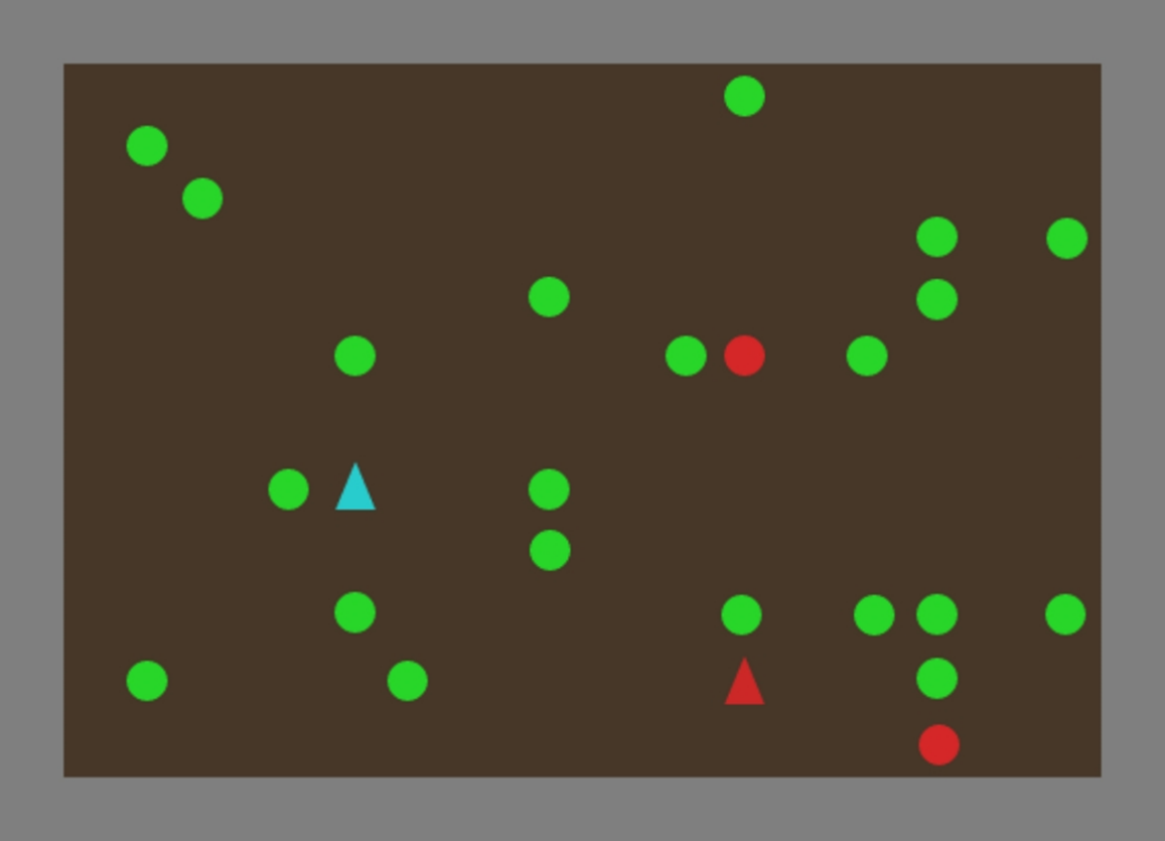} & 
\includegraphics[width=0.31\linewidth]{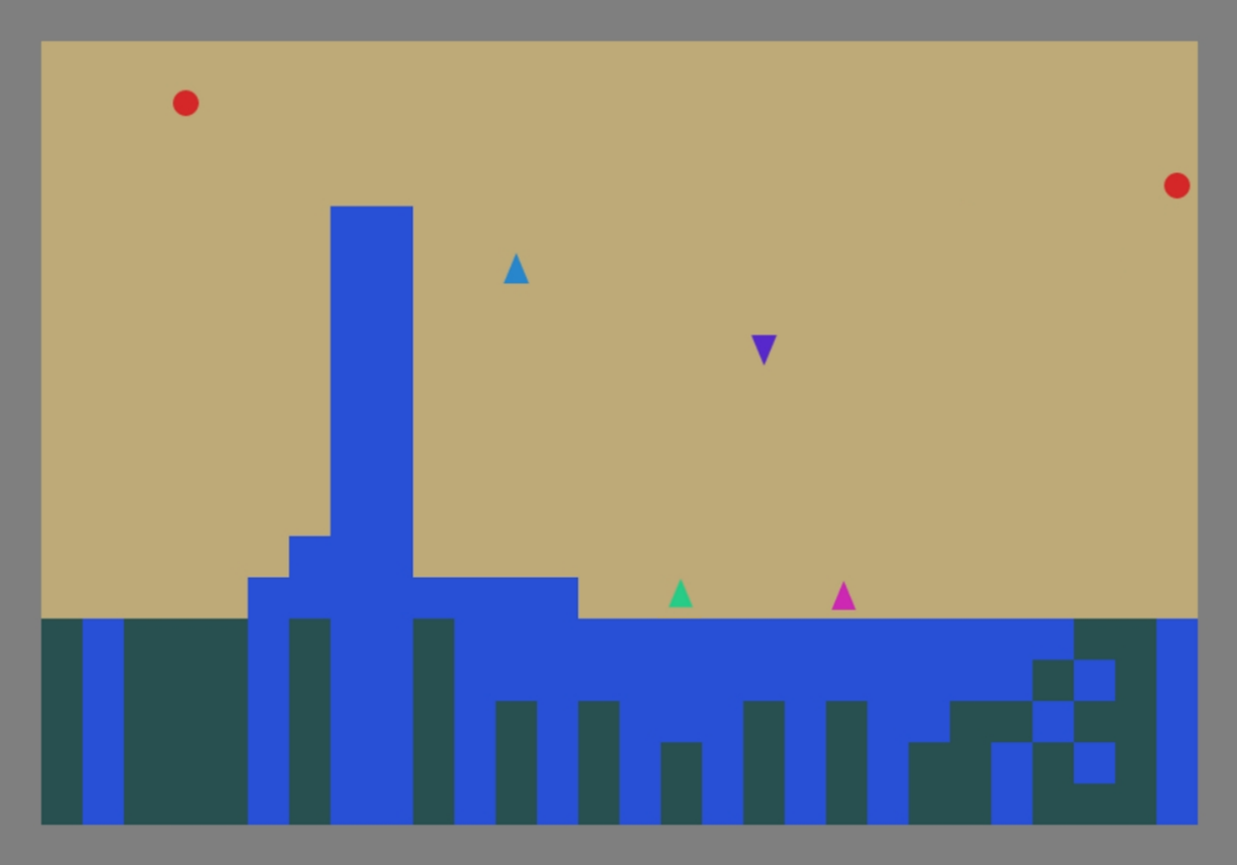} &
\includegraphics[width=0.29\linewidth]{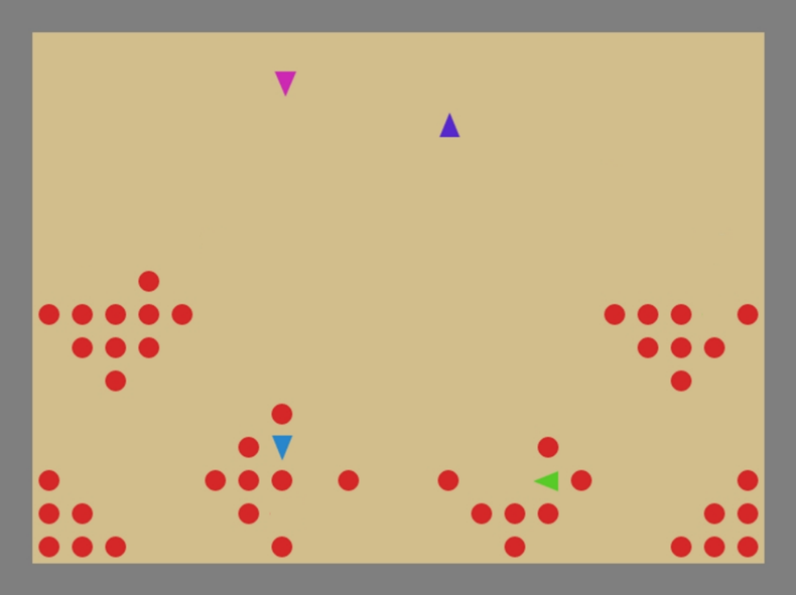} \\
(a) Unfair coins  &  (b) Cleanup & (c) Harvest
\end{tabular}
\caption{The environments considered in our experiments: (a) Unfair Coins---green coins appear more frequently than red coins, inducing fairness challenges; (b) Cleanup with distinct spawn positions---two agents (cyan and pink) spawn near waste areas, while the rest (blue and purple) spawn farther away; and (c) Harvest with distinct spawn positions---two agents (blue and green) spawn near apple (red) regions, while the rest spawn farther away.}
\label{fig:envs}
\end{figure}

\section{Experimental Results}

\subsection{Experimental Setup}\label{sec:setup}

\textbf{Environments}~~We conduct our experiments using the JAX-based codebase and environments provided by the SocialJAX suite~\cite{guo2025socialjax}. We modify the existing environments---Coins, Cleanup, and Harvest---to incorporate a fairness perspective. Specifically, since Cleanup and Harvest already involve inherent fairness dilemmas, we introduce only minor changes by assigning distinct respawn positions to the agents. For the Coin Game, which originally focuses on the conflict between individual and collective objectives, we introduce asymmetry in the potential rewards that agents can obtain, creating a disparity in individual incentives. Fig.\ref{fig:envs} illustrates the considered environments. We provide detailed descriptions in the following sections.

\textbf{Metric}~~As our goal is to maximize returns while ensuring fairness, both performance and fairness metrics should be jointly considered for evaluation. We use \textit{$\alpha$-fairness}~\cite{mo2000fair} as the evaluation metric, where, given individual returns $(r_1,\cdots, r_N)$, the fairness utility is defined as
\begin{align}
    U_{\alpha}(r_1,\cdots, r_N) =
    \begin{cases}
    \sum_{i=1}^N \frac{r_i^{1 - \alpha}}{1 - \alpha}, & \text{if } \alpha \ne 1, \\
    \sum_{i=1}^N \log(r_i), & \text{if } \alpha = 1.
    \end{cases}
\end{align}
Notably, the fairness utility recovers several well-known objectives for specific values of $\alpha$: it corresponds to the collective return when $\alpha = 0$, the geometric mean of individual rewards—also known as Nash Social Welfare---when $\alpha = 1$, and the minimum individual reward when $\alpha \to \infty$. Thus, $\alpha = 0$ reflects no consideration of fairness, and as $\alpha$ increases, the evaluation increasingly prioritizes fairness over aggregate performance.  In summary, we consider the following three representative instances of $\alpha$-fairness return in our evaluation: (i) average return (\textbf{Mean}, $\alpha = 0$), (ii) geometric mean return (\textbf{GeoMean}, $\alpha = 1$), and (iii) minimum individual return (\textbf{Min}, $\alpha \to \infty$).
Note that \textit{$\alpha$-fairness return considers both performance and fairness}, where $\alpha$ determines the trade-off between them. The reported results are averaged over four random seeds.

\textbf{Baselines}~~We evaluate FCGrad with six baselines: (a) collective reward optimization (Col), (b) individual reward optimization (Ind), (c) inequity aversion reward restructuring (IA) ~\cite{hughes2018inequity}, (d) weighted gradient combination of $g_{\text{ind}}$ and $g_{\text{col}}$ (denoted as Weighted), which corresponds to FCGrad without conflict handling, (e) PCGrad~\cite{yu2020gradient}, and (f) Altruistic Gradient Adjustment (AgA)~\cite{lialigning}. Note that baselines (d)-(f) use the same architecture as FCGrad, where two separate value functions are trained for individual and collective objectives; they differ only in the policy update rule based on $g_{\text{ind}}$ and $g_{\text{col}}$. All methods are implemented on top of the IPPO~\cite{de2020independent}.

\textbf{Hyperparameter}~~  We introduce a hyperparameter $\beta$ for FCGrad, which determines the weight between the collective and individual objectives when there is no conflict. $\beta$ plays a particularly important role in tasks that require high-level cooperation. We set $\beta$ to 0.5, 0.7, and 0.8 for the Unfair Coin Game, Cleanup, and Harvest, respectively. The same values of $\beta$ are used for the baseline method, Weighted. Additional hyperparameters for IPPO are provided in Appendix B.

\subsection{Unfair Coins} 

The Coins environment~\cite{lerer2017maintaining} consists of two agents (green and red) and two types of coins, each associated with one of the agents. When a coin appears, it is assigned a color with probabilities $p_{\text{green}}$ and $p_{\text{red}}$. An agent receives a reward of $1$ for collecting any coin, regardless of its color. However, collecting a coin of the opposite color imposes a penalty of $-2$ on the other agent, creating a conflict between individual gain and cooperative behavior. In contrast to the original setting~\cite{lerer2017maintaining}, where $p_{\text{green}}$ and $p_{\text{red}}$ are both set to 0.5---so that collecting coins matching each agent’s color naturally aligns with fairness and also maximizes the collective reward---we consider an unfair variant where $p_{\text{green}} = 15/16$ and $p_{\text{red}} = 1/16$, introducing an inherent asymmetry in coin appearances.
Although optimal collective performance still requires agents to collect coins matching their own color, this setup raises a fairness concern: the green agent receives substantially more rewards due to the higher frequency of green coins. To mitigate this imbalance and achieve a fairer outcome, the green agent must occasionally yield coins to the red agent, sacrificing some collective reward in favor of equity.

\begin{wrapfigure}{r}{0.5\textwidth}
\vspace{-4ex} %
\centering
\includegraphics[width=1\linewidth]{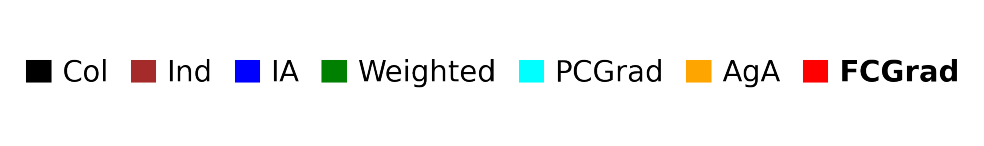} \\ \vspace{-2ex}
\hspace{-2ex}\includegraphics[width=0.45\textwidth]{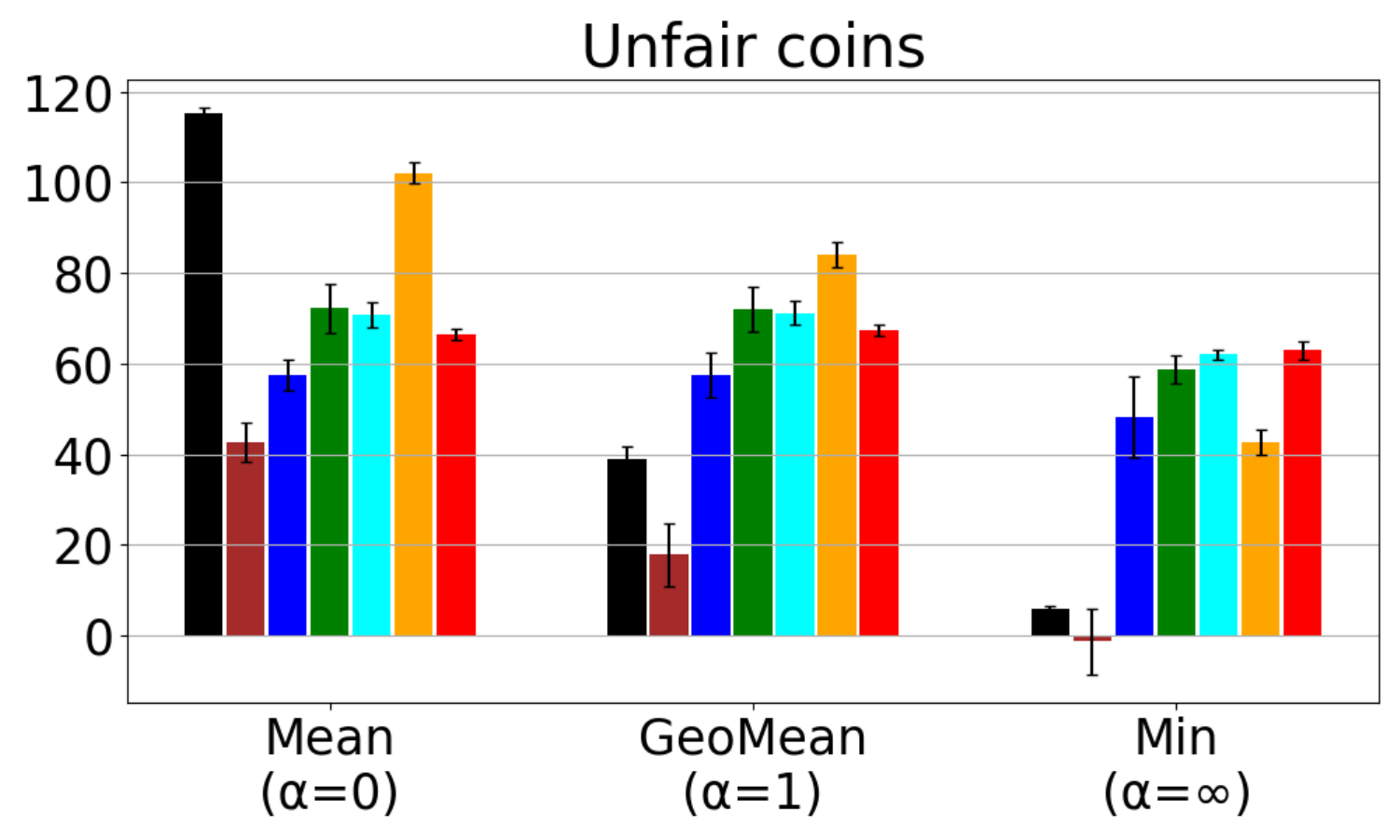}
\vspace{0.3em}
\begin{minipage}{0.24\textwidth}
    \includegraphics[width=\linewidth]{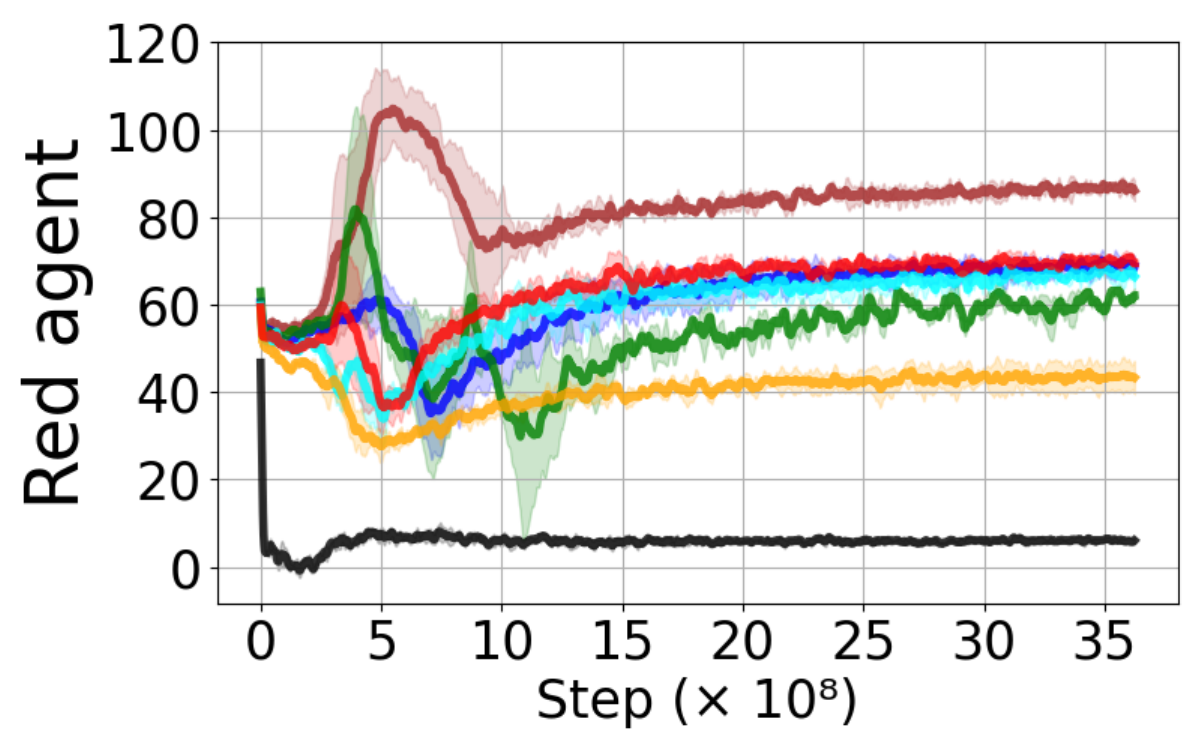} 
\end{minipage}%
\hfill
\begin{minipage}{0.24\textwidth}
    \includegraphics[width=\linewidth]{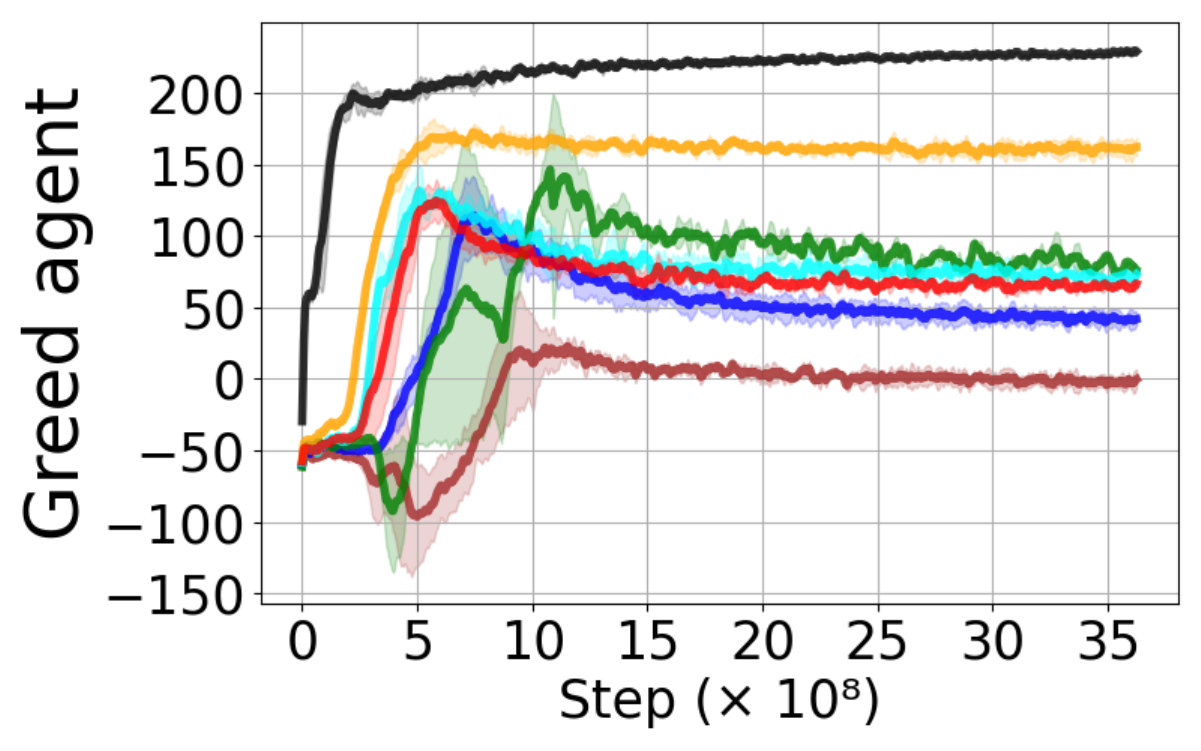}
\end{minipage}
\caption{Top: Performance across agents in terms of mean, geometric mean, and minimum. Note that a higher value of $\alpha$ places more emphasis on fairness. Bottom: Agent-wise returns---Red and Green Agents.}\vspace{-3ex}
\label{fig:result_coin}
\end{wrapfigure}

\textbf{Results.}~~In the Unfair Coin environment, achieving fairness requires the green agent to yield some of its coins to the red agent, thereby reducing its own reward. In other words, there exists a strong trade-off between collective performance and fairness. Therefore, we particularly focus on the performance trend with respect to $\alpha$, as well as the \textbf{Min} performance, which places greater emphasis on fairness---the return of the most disadvantaged agent.

Fig.~\ref{fig:result_coin} presents the $\alpha$-fairness returns in the unfair coin environment (top) and the individual return of the green and red agents (bottom). The performance of Col, Ind, and AgA is observed to decrease more dramatically as $\alpha$ increases compared to FCGrad and PCGrad, which are conflict-aware methods. Interestingly, both the collective and individual approaches result in extremely unfair outcomes, but in opposite directions. Col, which maximizes collective reward, trains both agents to collect their own coins. As a result, the green agent, with more coin opportunities, gains higher returns, leading to unfairness. In contrast, with the individual objective, the red agent outperforms the green agent, possibly because the green agent is more frequently penalized by negative rewards due to the abundance of green coins. Meanwhile, the red agent learns without such penalties, accelerating its progress. 
However, FCGrad shows little variation across agents as $\alpha$ changes, indicating achieved fairness. Notably, FCGrad outperforms the baselines in terms of \textbf{Min} performance. As shown in Fig.~\ref{fig:result_coin}, both the green and red agents converge to nearly identical returns, showing that fairness is effectively achieved.

\subsection{Cleanup} 

The Cleanup environment consists of $N=4$ agents, apples, and waste. Each agent receives a reward of $1$ for collecting an apple. Apples grow in an orchard, but their growth depends on the amount of waste present in the environment. Waste accumulates at a constant rate, and beyond a certain threshold, apple growth ceases entirely. Therefore, in order to sustain apple regrowth, some agents must sacrifice their immediate reward by cleaning up the waste. This creates a social dilemma, as the necessary act of cleaning benefits the group but does not provide direct individual reward, thereby generating a tension between self-interest and cooperative behavior. In contrast to the original configuration, where agents are randomly spawned across the map, we fix the spawn positions of agents: some (Agents 2 and 3 in our case) are placed near the apple orchard, while others (Agents 0 and 1) are positioned closer to the waste area. This spatial asymmetry further amplifies the conflict between fairness and efficiency. Note that, unlike the Unfair Coin, Cleanup introduces an intertemporal perspective, involving a trade-off between short-term individual interest and long-term collective interest~\cite{hughes2018inequity}.

\begin{figure}[t]
\centering
\includegraphics[width=0.6\linewidth]{figures/legend_only.pdf} \\ \vspace{-2ex}
\begin{minipage}[t]{0.48\textwidth}
    \centering
    \includegraphics[width=0.9\linewidth]{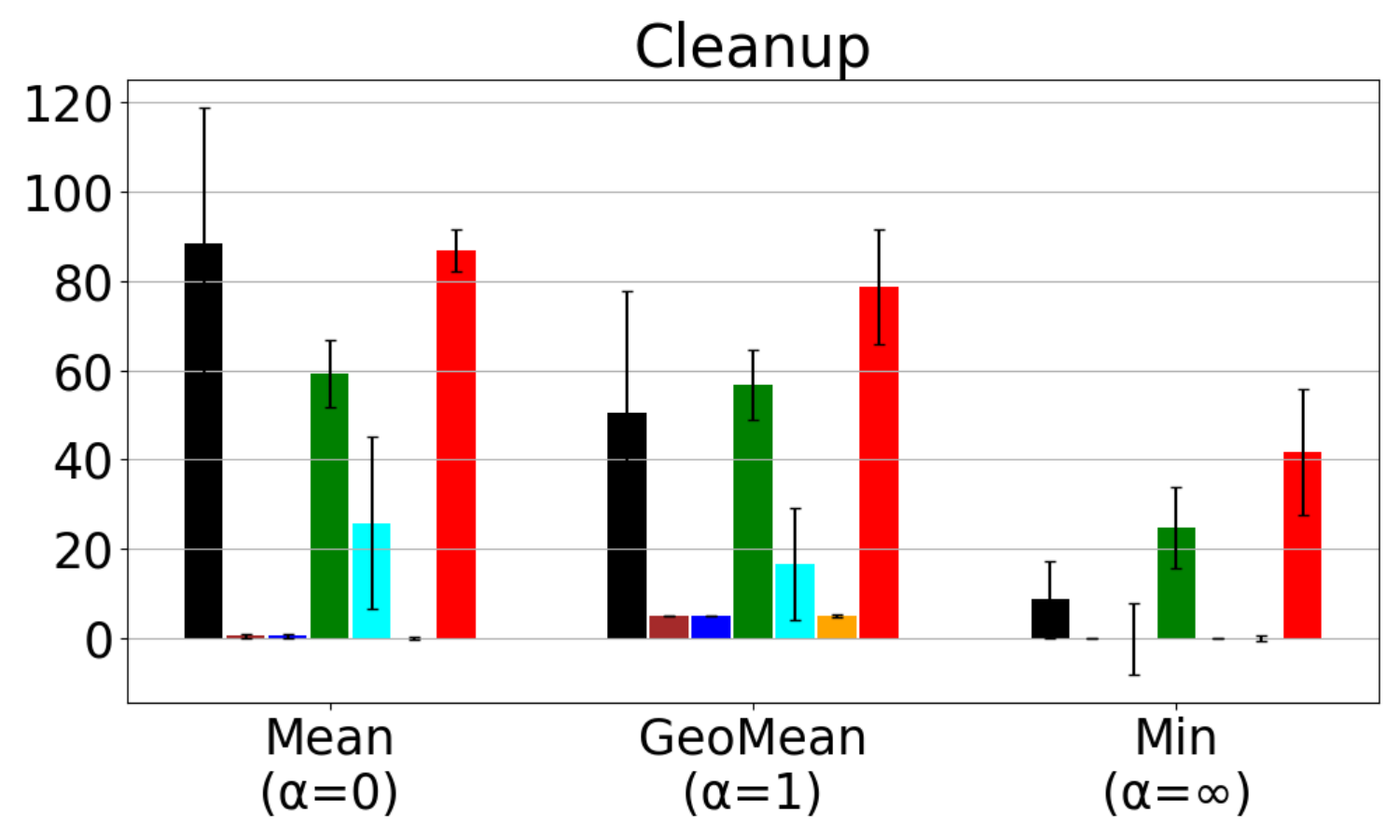}
    \caption*{(a) $\alpha$-fairness returns (Cleanup)}
\end{minipage}%
\hfill
\begin{minipage}[t]{0.5\textwidth}\vspace{-23ex}
    \centering
    \begin{minipage}[t]{0.48\linewidth}
        \includegraphics[width=\linewidth]{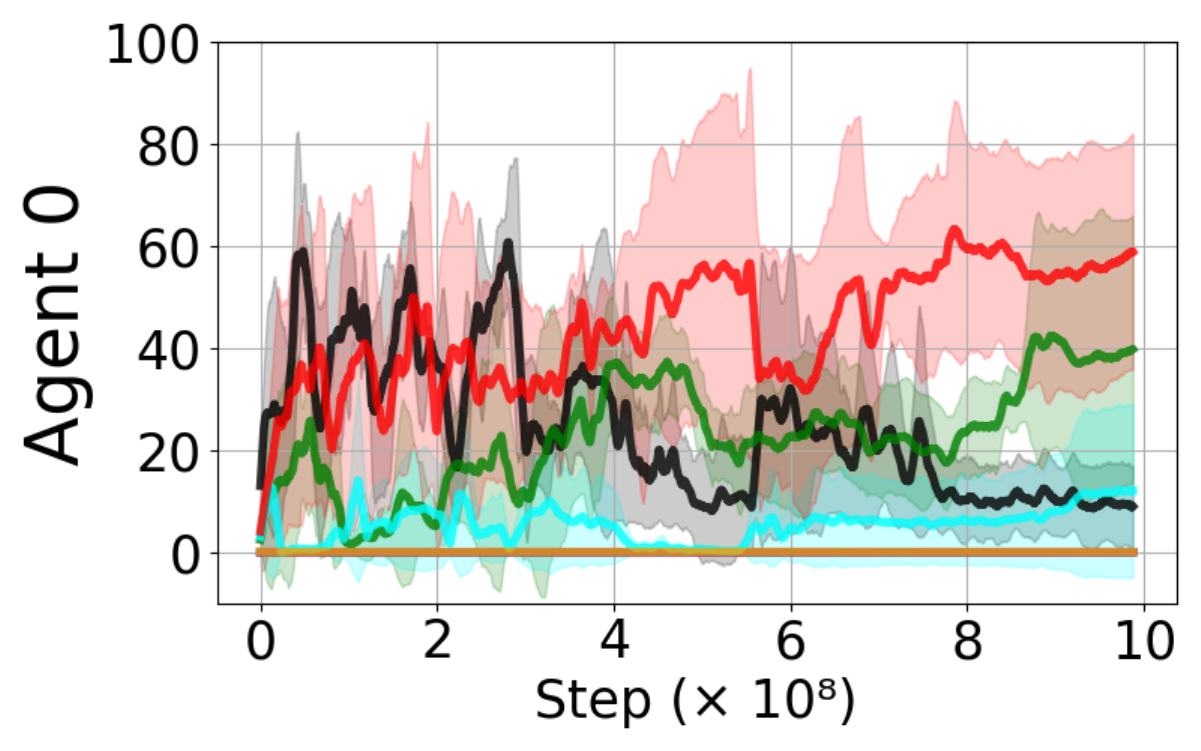}
    \end{minipage}%
    \hfill
    \begin{minipage}[t]{0.48\linewidth}
        \includegraphics[width=\linewidth]{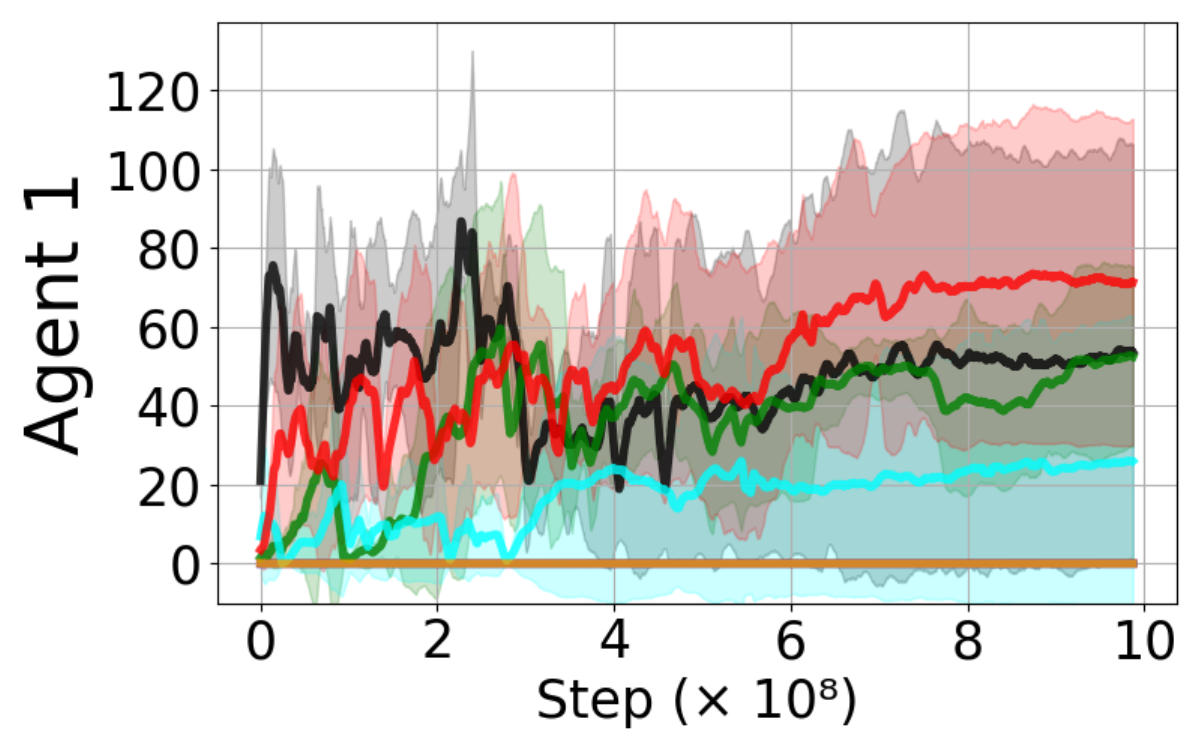}
    \end{minipage}
    
    \begin{minipage}[t]{0.48\linewidth}
        \includegraphics[width=\linewidth]{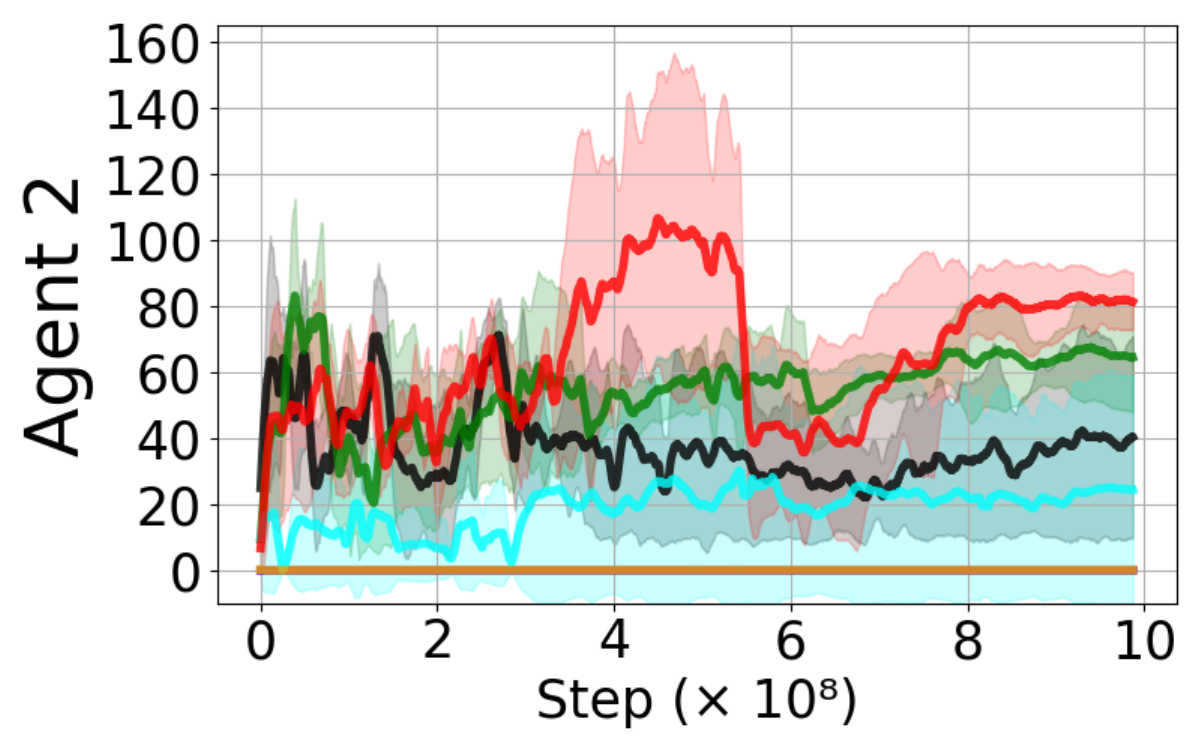}
    \end{minipage}%
    \hfill
    \begin{minipage}[t]{0.48\linewidth}
        \includegraphics[width=\linewidth]{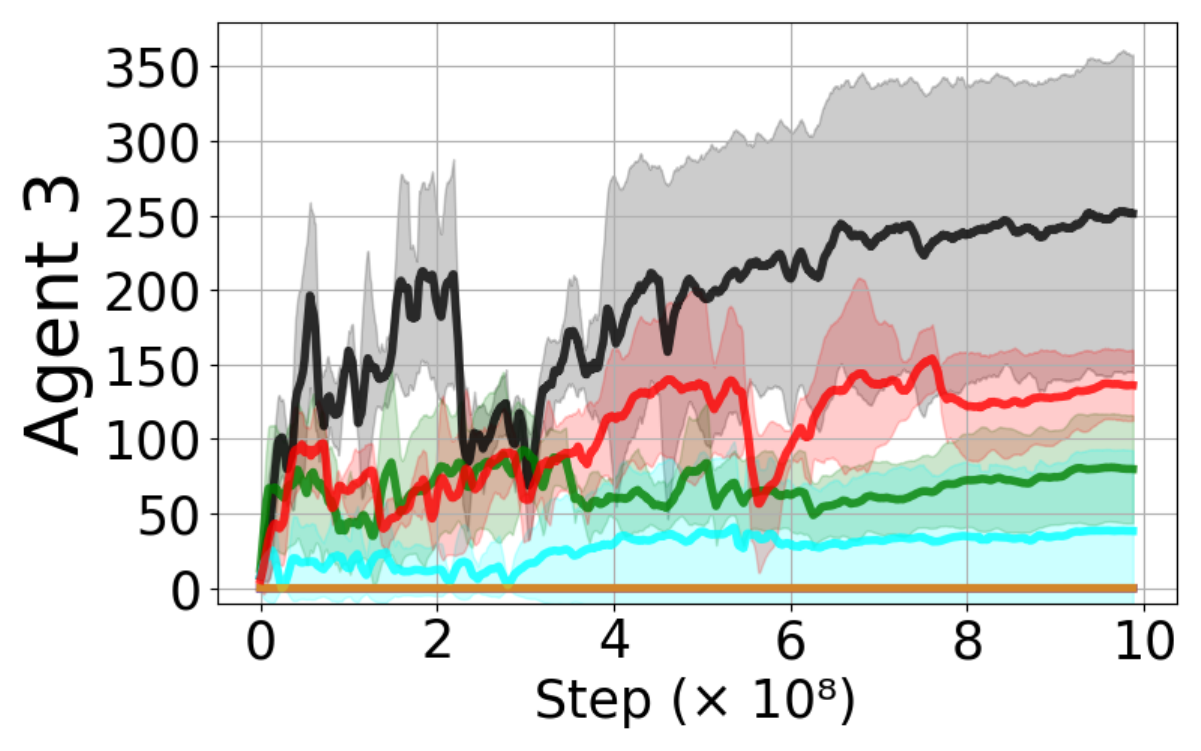}
    \end{minipage}
    \vspace{-1ex}
    \caption*{(b) Individual returns (Cleanup)}
\end{minipage}

\vspace{0ex}

\begin{minipage}[t]{0.48\textwidth}\vspace{-12.5ex}
    \centering
    \includegraphics[width=0.9\linewidth]{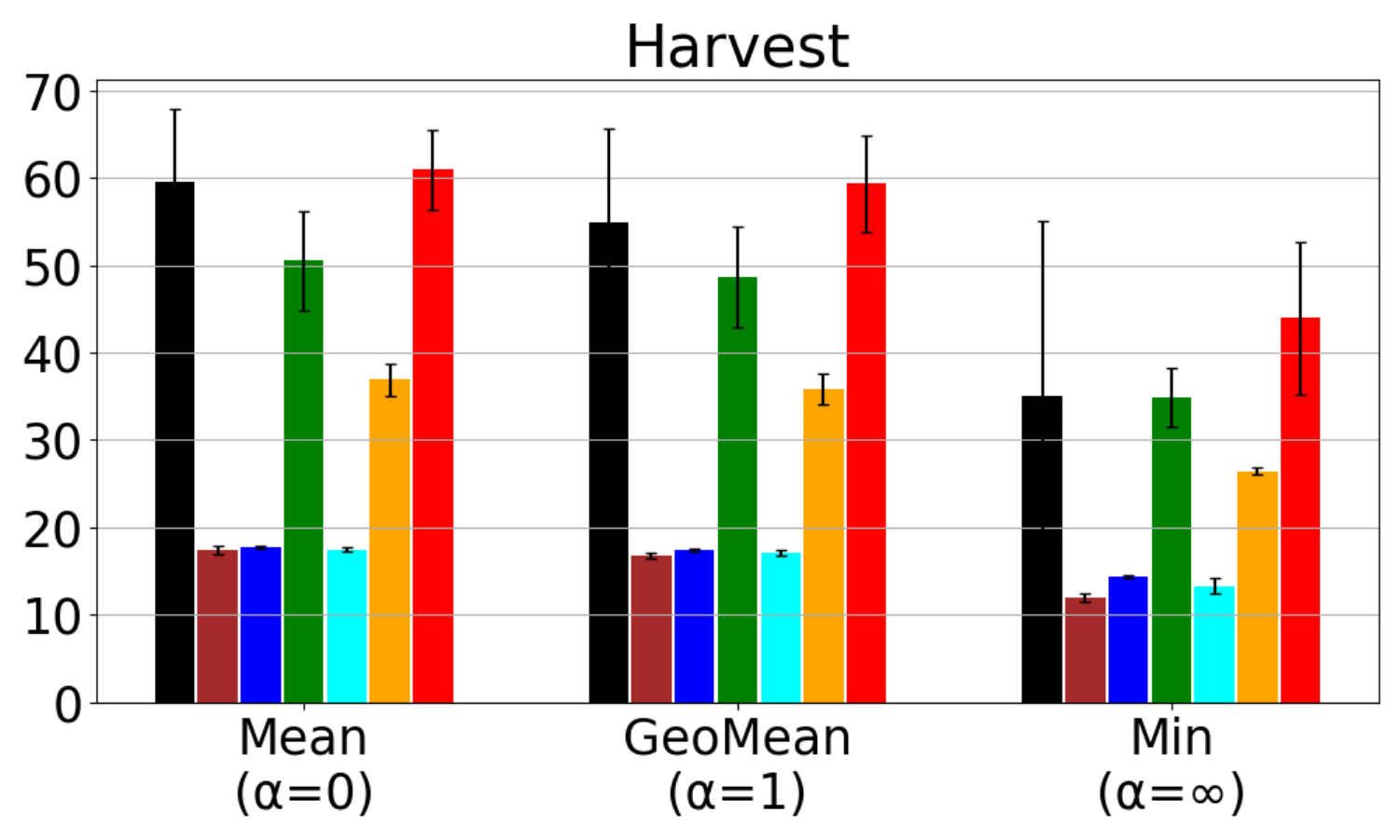}
    \caption*{(c) $\alpha$-fairness returns  (Harvest)}
\end{minipage}%
\hfill
\begin{minipage}[t]{0.5\textwidth}
    \centering
    \begin{minipage}[t]{0.48\linewidth}
        \includegraphics[width=\linewidth]{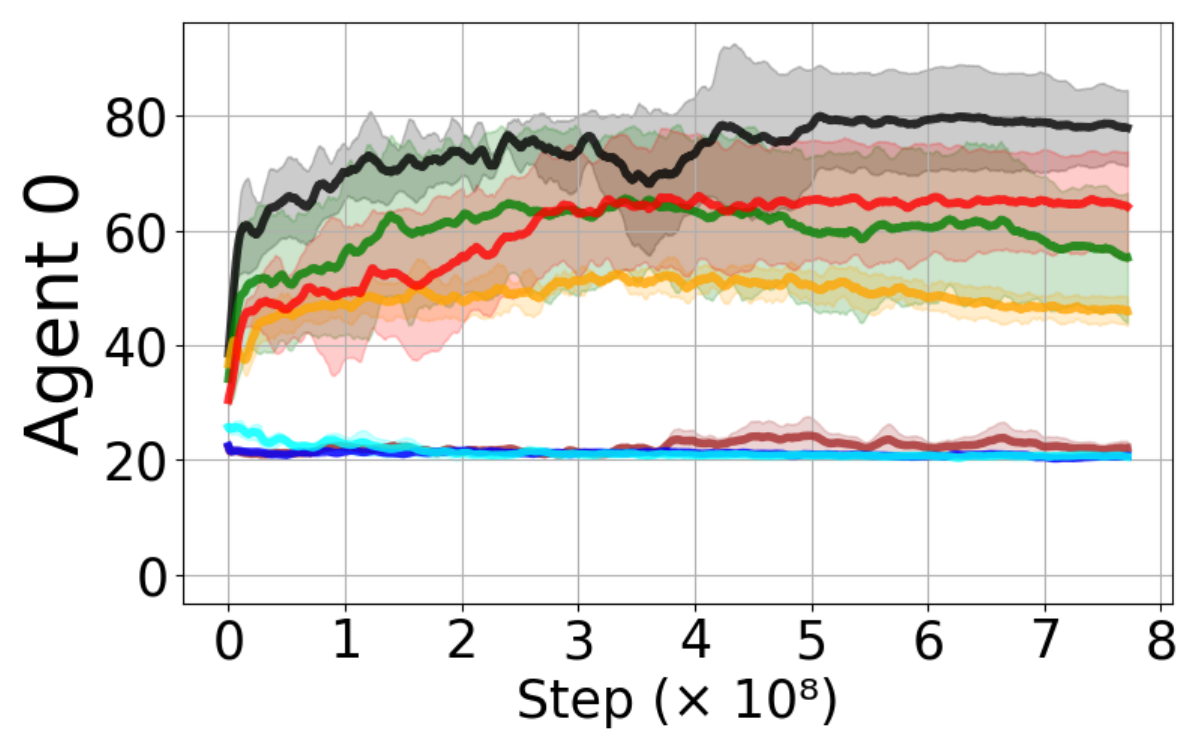}
    \end{minipage}%
    \hfill
    \begin{minipage}[t]{0.48\linewidth}
        \includegraphics[width=\linewidth]{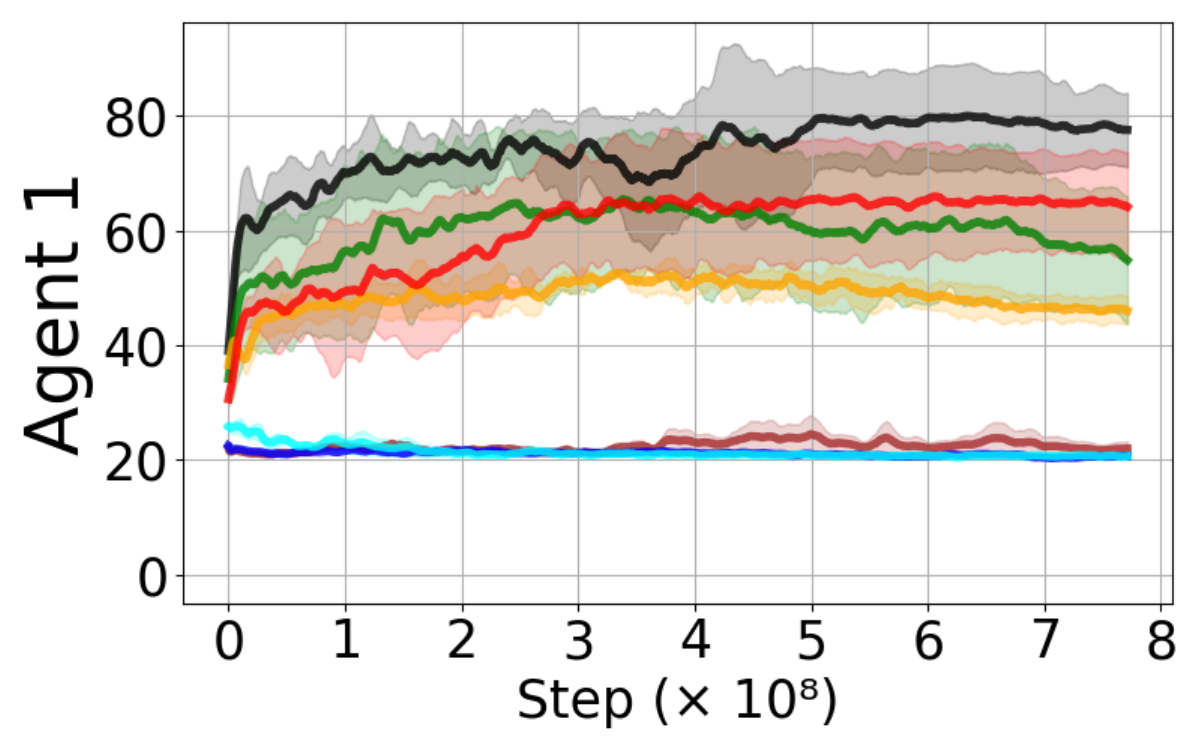}
    \end{minipage}
    \vspace{1ex}
    
    \begin{minipage}[t]{0.48\linewidth}
        \includegraphics[width=\linewidth]{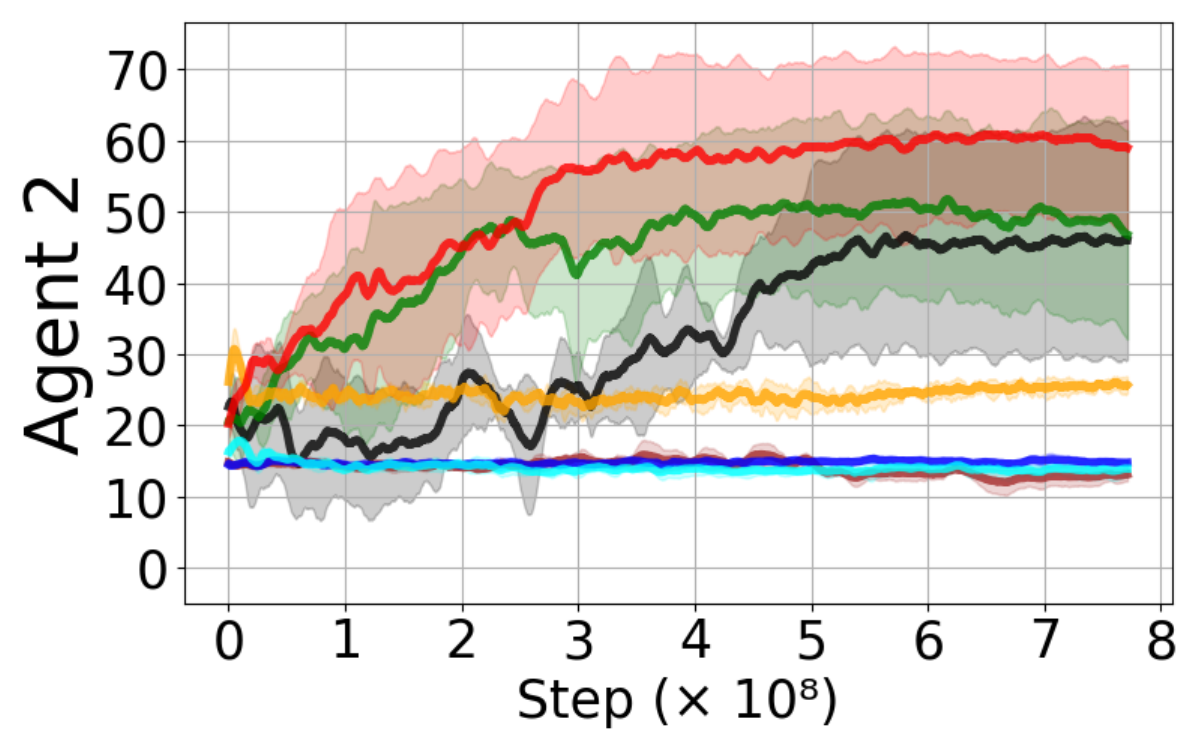}
    \end{minipage}%
    \hfill
    \begin{minipage}[t]{0.48\linewidth}
        \includegraphics[width=\linewidth]{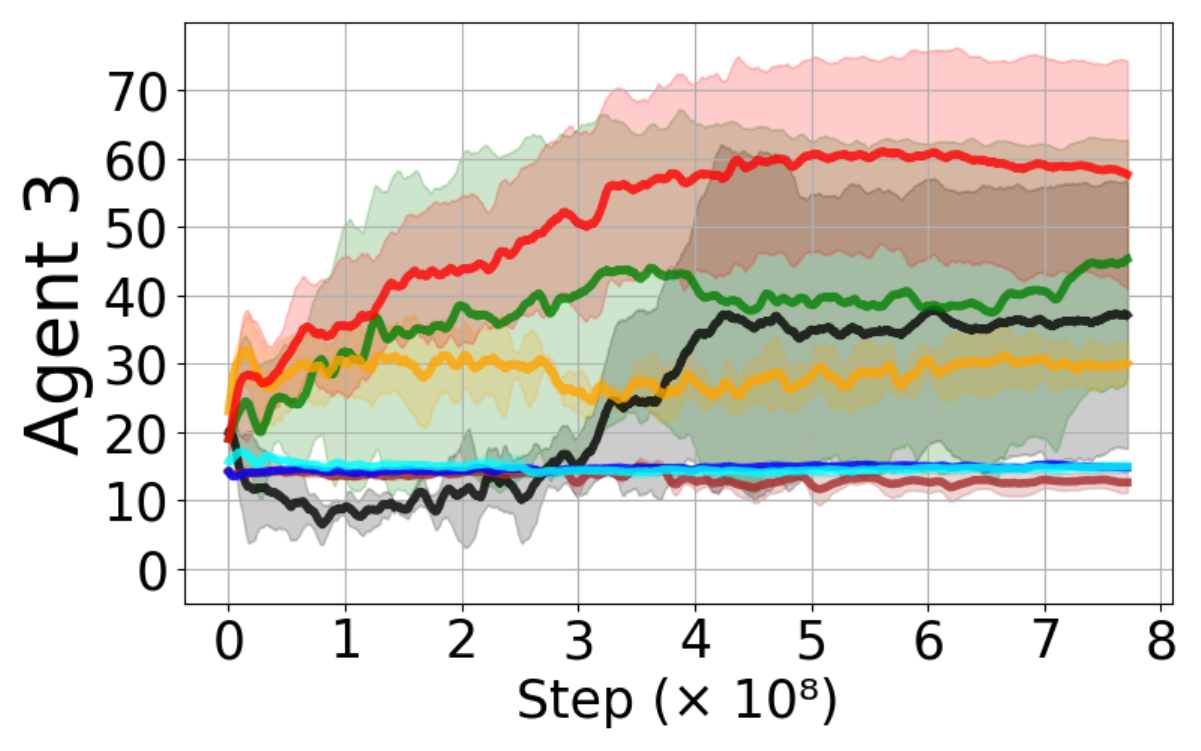}
    \end{minipage}
    \caption*{(d) Individual returns (Harvest)}
\end{minipage}
\caption{$\alpha$-fairness returns and individual returns in the cleanup and harvest environments.}\vspace{-2ex}
\label{fig:result_twoenv}
\end{figure}

\textbf{Results.} ~~Fig.~\ref{fig:result_twoenv} (a) and (b) show the $\alpha$-fairness performance and individual rewards during training in the Cleanup environment. FCGrad outperforms the baselines in terms of both \textbf{GeoMean} and \textbf{Min}, which reflect not only total return but also fairness. In addition, FCGrad achieves comparable performance to Col in terms of \textbf{Mean}, which is the optimization target of Col. As shown in Fig.~\ref{fig:result_twoenv} (b), under Col, Agent 3 learns to monopolize apple collection, while Agent 0 is trained to sacrifice by primarily cleaning waste. In contrast, FCGrad leads all four agents to obtain reasonably similar returns---demonstrating more fair behavior and achieving the best result in terms of \textbf{Min}. Since using the collective reward is essential in this environment, methods that rely heavily on individual rewards, such as Ind and IA, fail to learn effectively. In addition, AgA fails to properly balance between individual and collective objectives, also struggle to learn successfully.


\subsection{Harvest}

The Harvest environment features $N=4$ agents and apples distributed across orchard patches. Each agent receives a reward of 1 per apple, but regrowth is stochastic and depends on nearby apples within a fixed radius. Over-harvesting depletes resources, risking environmental collapse, and thus agents must coordinate implicitly to sustain long-term returns. This creates a social dilemma between short-term individual gain and long-term collective benefit. We also introduce spatial asymmetry: Agents 0 and 1 spawn near apples, while Agents 2 and 3 spawn farther away, making collection easier for the former. Similar to the Cleanup, Harvest also poses intertemporal challenges for both cooperation and fairness.

\begin{wrapfigure}{r}{0.4\textwidth} %
\centering
\vspace{-0ex}
\begin{tabular}{c}
\includegraphics[width=0.85\linewidth]{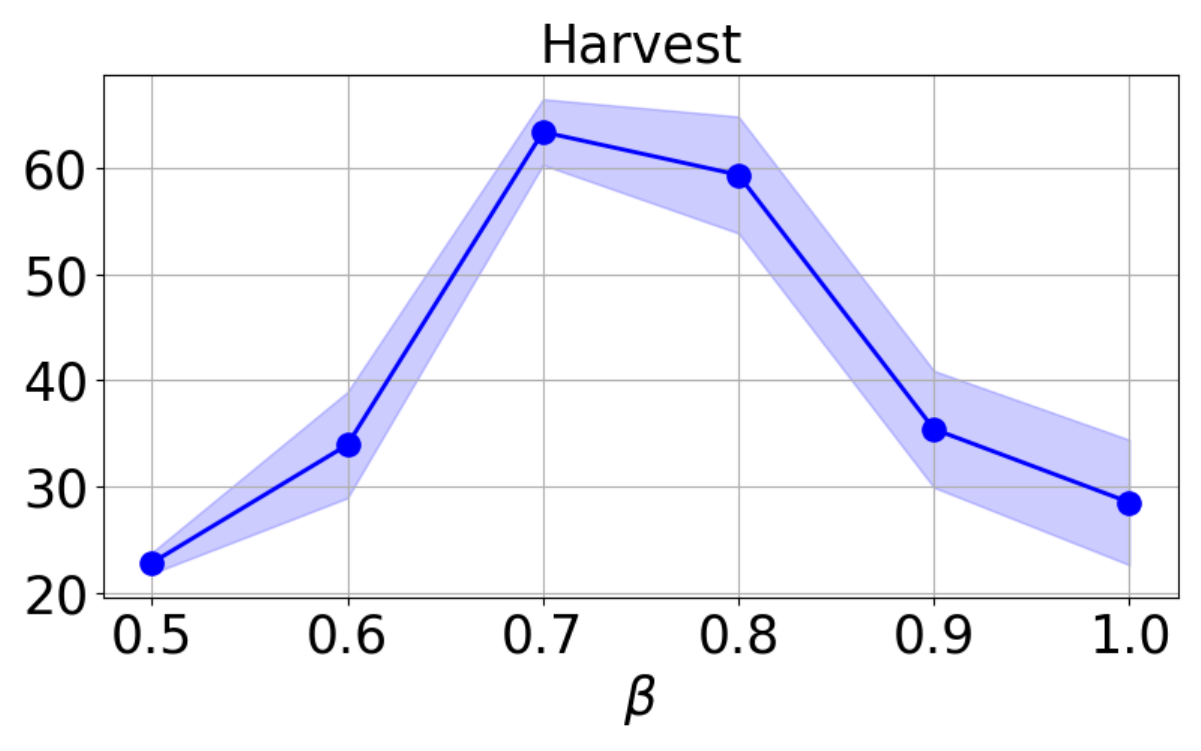} \\
\end{tabular}
\caption{GeoMean of FCGrad with respect to $\beta$ in the harvest environment.}
\label{fig:abalation}
\vspace{-1ex}
\end{wrapfigure}

\textbf{Results.}~~ Fig.~\ref{fig:result_twoenv} (c) and (d) show the $\alpha$-fairness returns and individual agent returns during training. FCGrad outperforms the baselines across the considered $\alpha$ values. With the Col, Agents 0 and 1 achieve higher returns than Agents 2 and 3, indicating that they focus solely on collecting apples while accounting for the intertemporal dilemma, but not addressing the resulting unfairness toward Agents 2 and 3. In contrast, FCGrad leads all four agents to achieve similar returns, implying that Agents 0 and 1 take into account the outcomes of Agents 2 and 3. Similar to the results in Cleanup, methods that rely heavily on individual rewards, such as Ind and IA, perform poorly, though they achieve marginal learning. AgA performs better than the individual-reward-based methods, but still underperforms compared to FCGrad.


\begin{wraptable}{r}{0.5\textwidth}  %
\centering
\vspace{-4ex}
\footnotesize
\setlength{\tabcolsep}{3pt}
\renewcommand{\arraystretch}{1.1}
\begin{tabular}{lcccccc}
\toprule
\textbf{} & \multicolumn{2}{c}{\textbf{Coin}} & \multicolumn{2}{c}{\textbf{Cleanup}} & \multicolumn{2}{c}{\textbf{Harvest}} \\
\cmidrule(lr){2-3} \cmidrule(lr){4-5} \cmidrule(lr){6-7}
\textbf{Alg} & Gini & Jain & Gini & Jain & Gini & Jain \\
\midrule
Col & 0.474 & 0.526 & 0.558 & 0.432 & 0.182 & 0.882 \\
Ind & 0.509 & 0.498 & 0.522 & 0.515 & 0.136 & 0.936 \\
IA & 0.122 & 0.942 & 0.536 & 0.497 & \bf{0.087} & \bf{0.973} \\
Weighted & 0.048 & 0.991 & \bf{0.266} & \bf{0.801} & 0.146 & 0.928 \\
PCGrad & \bf{0.039} & \bf{0.994} & 0.469 & 0.572 & 0.101 & \bf{0.965} \\
AgA & 0.238 & 0.749 & 0.331 & 0.723 & 0.123 & 0.948 \\
FCGrad & \bf{0.010} & \bf{0.999} & \bf{0.223} & \bf{0.835} & \bf{0.093} & 0.959 \\
\bottomrule
\end{tabular}
\caption{Addtional fairness evaluation using Gini coefficient and Jain's index. Lower Gini and higher Jain indicate greater fairness. Top-2 most fair scores in each column are highlighted in bold.}
\label{tab:fairness}\vspace{-3ex}
\end{wraptable}

\subsection{Additional Analysis: Ablation and Fairness Metrics}

\textbf{Weighting factor:} $\beta$ determines the balance between the collective and individual objectives when no conflict is detected. It plays a particularly important role in tasks that require high-level cooperation. For example, in Cleanup, ignoring the collective objective makes it difficult for agents to discover how to improve their individual rewards. We observed this phenomenon in the previous section---solely maximizing individual rewards does not perform well. We present the \textbf{GeoMean} performance of FCGrad in the Harvest environment in Fig.~\ref{fig:abalation}, which shows that a $\beta$ value between 0.7 and 0.8 yields the best performance. Thus, $\beta$ reflects the required degree of cooperation over self-interest.

\textbf{Additional Fairness metrics: } We additionally evaluate fairness using the Gini coefficient~\cite{david1968miscellanea} and Jain's index~\cite{jain1984quantitative}. The Gini coefficient is defined as $\text{Gini}(r_1,\cdots, r_N) = \frac{\sum_{i=1}^{N} \sum_{j=1}^{N} |r_i - r_j|}{2N \sum_{i=1}^{N} r_i}$ and Jain's index is defined as $\text{Jain}(r_1,\cdots, r_N) = \frac{\left( \sum_{i=1}^{N} r_i \right)^2}{N \sum_{i=1}^{N} r_i^2},$ where both metrics range between 0 and 1 and lower Gini and higher Jain values indicate better fairness. Table~\ref{tab:fairness} presents the results, showing that FCGrad generally achieves superior fairness.

\section{Conclusion}

In this work, we address the long-standing challenge of achieving both cooperation and fairness in mixed-motive multi-agent RL. We propose FCGrad, a conflict-aware gradient adjustment method that explicitly resolves gradient-level conflicts between individual and collective objectives. FCGrad dynamically adjusts the update direction based on which objective is more disadvantaged by projecting one gradient onto the normal plane of the other. We theoretically prove that this mechanism guarantees monotonic improvement and convergence of both objectives to the same value. Consequently, individual objectives across agents also converge, ensuring fairness. Extensive experiments in the Unfair Coin environment and sequential social dilemma settings, Cleanup and Harvest, demonstrate that FCGrad not only improves overall performance but also achieves superior fairness, as measured by $\alpha$-fairness return metrics.

\textbf{Limitation}~~In practice, the recurrence of gradient conflicts, required for our theoretical guarantee, may not hold, as it can be influenced by the weighting factor in non-conflict cases. Understanding this interplay is a promising direction for future work.

\textbf{Broader Impact}~~Our work promotes fairness in learned behaviors, potentially preventing emergent inequalities in decentralized systems. We believe it has a positive societal impact.

\section{Acknowledgement}
This work was supported by the ONR MURI grant N00014-25-1-2116.

\bibliographystyle{plain} 
\bibliography{refs}

\newpage

\appendix

\section{Theoretical Results}

\begin{lemma} Let \( J: \mathbb{R}^d \to \mathbb{R} \) be a continuously differentiable and \( L \)-smooth function.  
Let \( g_1 = \nabla_\theta J(\theta) \) be the gradient of \( J \) at point \( \theta \), and let \( g_2 \in \mathbb{R}^d \) be any vector satisfying \( \langle g_1, g_2 \rangle > 0 \).  
Then, for small step size $\eta< \frac{2 \langle g_1, g_2 \rangle}{L \|g_2\|^2}$,  the update \( \theta \leftarrow \theta + \eta g_2 \) yields a strict improvement:
\[
J(\theta + \eta g_2) > J(\theta).
\]
\end{lemma}

\textit{Proof}. Since \( J \) is \( L \)-smooth, for any \( \theta \in \mathbb{R}^d \), update direction \( g_2 \in \mathbb{R}^d \), and step size \( \eta > 0 \), the following inequality holds:
\[
J(\theta + \eta g_2) \geq J(\theta) + \eta \langle \nabla_\theta J(\theta), g_2 \rangle - \frac{L}{2} \eta^2 \|g_2\|^2.
\]
Let \( g_1 = \nabla_\theta J(\theta) \). Then:
\[
J(\theta + \eta g_2) \geq J(\theta) + \eta \langle g_1, g_2 \rangle - \frac{L}{2} \eta^2 \|g_2\|^2.
\]
Define the right-hand side as a function of \( \eta \):
\[
\Delta(\eta) := \eta \langle g_1, g_2 \rangle - \frac{L}{2} \eta^2 \|g_2\|^2.
\]
Since \( \langle g_1, g_2 \rangle > 0 \), this is a concave quadratic function that is positive for small enough \( \eta \).  
Specifically, the inequality \( \Delta(\eta) > 0 \) holds when:
\[
\eta < \frac{2 \langle g_1, g_2 \rangle}{L \|g_2\|^2}.
\]
Therefore, for any \( \eta \in \left(0,\ \frac{2 \langle g_1, g_2 \rangle}{L \|g_2\|^2} \right) \), we have:
\[
J(\theta + \eta g_2) > J(\theta).
\]

\begin{theorem}
\label{theorem:increase_appendix}
Assume $V_{\text{ind}}(\theta)$ and $V_{\text{col}}(\theta)$ are differentiable and L-smooth. Let the update direction $g$ be defined as in Equation~\ref{eq:ourgradient}. Then, for a sufficiently small step size $\eta>0$, the update $\theta \leftarrow \theta + \eta g$ yields monotonically non-decreasing improvements in both $V_{\text{col}}(\theta)$ and $V_{\text{int}}(\theta)$.
\end{theorem}

We consider three cases:

\textbf{Case 1:} (Non-conflict) \( g_{\text{ind}} \cdot g_{\text{col}} \geq 0 \). Then \( g = \beta g_{\text{ind}} + (1 - \beta) g_{\text{col}} \). Since \( g_{\text{ind}}, g_{\text{col}} \) are ascent directions for \( V_{\text{ind}}, V_{\text{col}} \), respectively, their convex combination also satisfies:
\begin{align}
    & g_{\text{ind}}\cdot g = \beta \|g_{\text{ind}}\|^2 + (1-\beta)  g_{\text{ind}} \cdot g_{\text{col}} > 0 \\
    & g_{\text{col}}\cdot g = \beta g_{\text{col}} \cdot g_{\text{ind}} + (1-\beta) \|g_{\text{col}}\|^2  > 0
\end{align}

Since $g_{\text{ind}}\cdot g$ and $g_{\text{ind}}\cdot g$ are positive, according to Lemma 3.1, $g$ yields a strict improvement in both $V_{\text{ind}}$ and $V_{\text{col}}$.

\textbf{Case 2:} (Conflict) \( g_{\text{ind}} \cdot g_{\text{col}} < 0 \) and \( V_{\text{ind}}(\theta) < V_{\text{col}}(\theta) \).
We then use: \( g = g_{\text{ind}} - \frac{g_{col} \cdot g_{\text{ind}}}{\|g_{col}\|^2}g_{col} \).
Now,
\begin{align}
    & g_{\text{ind}} \cdot g = g_{\text{ind}} \cdot g_{\text{ind}} - \frac{ (g_{\text{ind}} \cdot g_{\text{col}})}{\|g_{\text{col}}\|^2} (g_{\text{ind}} \cdot g_{\text{col}}) = \frac{\|g_{\text{ind}}\|^2\|g_{\text{col}}\|^2 -  (g_{\text{ind}}\cdot g_{\text{col}})^2}{\|g_{\text{col}}\|^2} > 0 \nonumber \\
    & g_{\text{col}} \cdot g = g_{\text{col}} \cdot g_{\text{ind}} - \frac{(g_{\text{ind}} \cdot g_{\text{col}})}{\|g_{\text{col}}\|^2} \langle g_{\text{col}}, g_{\text{col}}\rangle = 0
\end{align}

Since $g_{\text{ind}}\cdot g$ is positive,  according to Lemma 3.1, $g$ yields a strict improvement in $V_{\text{ind}}$. In addition, since $g_{\text{col}}\cdot g$ is zero, $g$ does not decrease $V_{\text{col}}$.

\textbf{Case 3:} (Conflict) \( g_{\text{ind}} \cdot g_{\text{col}} < 0 \) and \( V_{\text{ind}}(\theta) > V_{\text{col}}(\theta) \). Symmetric to Case 2: $g$  yields a strict improvement in both $V_{\text{col}}$ and does not decrease $V_{\text{ind}}$.

Thus, in all cases, $g$ induces monotonically non-decreasing improvements in $V_{\text{ind}}$ and $V_{\text{col}}$.

\begin{lemma}[Single conflict step]
\label{lem:conflict}
When the conflict happens (i.e., $(g_{\text{ind}} \cdot g_{\text{col}}) < 0$), then for sufficiently small step size, $0\le \eta_t\le \|\delta_t\|/L$, we have 
\begin{align}
    L_{t+1}-L_t \leq
- \frac{\eta_t}{2}\,|\delta_t|\,\|d_t\|^{2}.
\end{align}
\end{lemma}

\textit{Proof}. When $\delta_t<0$ (i.e. $V_{\text{col}}>V_{\text{ind}})$, we use $g=g_{\text{ind}}-\frac{g_{\text{ind}} \cdot g_{\text{col}}}{\| g_{\text{col}}\|^2} g_{\text{col}}$. 
Since $L$ is L-smooth function, the following holds
\begin{align}
    L_{t+1} - L_t \leq \eta_t (\nabla L_t \cdot g) + \frac{L}{2} \eta_t^2 \|g\|^2
\end{align}
Here, 
\begin{align}
    (\nabla L_t \cdot g) = \delta_t (g_{\text{ind}} - g_{\text{col}}) \cdot g = \delta_t (g_{\text{ind}} \cdot g - g_{\text{col}}\cdot g) = \delta_t (g_{\text{ind}} \cdot g) = \delta_t \|g\|^2
\end{align}
Thus, we have
\begin{align}
    L_{t+1} - L_t \leq  \eta_t \delta_t \|g\|^2 + \frac{L}{2} \eta_t^2 \|g\|^2 \| \leq \eta_t \delta_t \|g\|^2 + \frac{\|\delta_t\|\eta_t}{2} \|g\|^2 = -\frac{\eta_t}{2}\|\delta_t\| \|g\|^2 
\end{align}

\begin{lemma}[Single non‑conflict step]
\label{lem:conflict}
When the conflict does not happen, (i.e., $(g_{\text{ind}} \cdot g_{\text{col}}) \geq 0$), the proposed gradient is used. We assume that the step size meets the Robbins-Monro conditions (i.e. $\sum_{t=0}^{\infty}\eta_t=\infty,~~
\sum_{t=0}^{\infty}\eta_t^{\,2}<\infty.$) Then, the following holds:
\begin{align}
    \sum_{t\in \mathcal{N}} \| L_{t+1} - L_t \| < \infty
\end{align}
where $\mathcal N$ is the set of all non‑conflict indices.
\end{lemma}

\textit{Proof}. $g = \beta g_{\text{ind}} + (1-\beta) g_{\text{col}}$. Let us define $G := \sup_{t}\bigl(\|g_{1,t}\|+\|g_{2,t}\|\bigr)\ (<\infty$).

Since $L$ is L-smooth, we have
\begin{align}
    L_{t+1}-L_t \leq \eta_t\langle\nabla_\theta L_t,\, g\rangle
+\frac{L}{2}\,\eta_t^{2}\|g\|^{2}.
\end{align}
Since $\nabla_{\theta}L_t=\delta_t (g_{\text{ind}} - g_{\text{col}})$, 
\begin{align}
    \|\langle\nabla_\theta L_t,\, g\rangle\| &= \|\delta_t \langle g_{\text{ind}} - g_{\text{col}}, \beta g_{\text{ind}} + (1-\beta) g_{\text{col}} \rangle \|\\
    &\leq |\delta_t| \Big[ \beta\|g_{\text{ind}}\|\|g_{\text{ind}}-g_{\text{col}}\| +(1-\beta) \|g_{\text{col}}\|\|g_{\text{ind}}-g_{\text{col}}\|  \Big]         ~~~ (\text{Cauchy-Schwarz}) \\
    & \leq |\delta_t| \Big[ \beta \|g_{\text{ind}}\| + (1-\beta) \|g_{\text{col}} \|    \Big] 2G \leq 2G^2 |\delta_t|.
\end{align}
Based on the assumption of the step size $\eta$ ($\eta_t \leq |\delta_t|/L$), we have
\begin{align}
    |\eta_t\langle\nabla_\theta L_t,d_t\rangle|
\;\le\;
2G^{2}\,|\delta_t|\,\eta_t
\;\le\;
2G^{2}L\,\eta_t^{2}.
\tag{C}
\end{align}

Since $\|g\| = \|\beta g_{\text{ind}} + g_{\text{col}}\| \leq \beta \| g_{\text{ind}} \| + (1-\beta) \| g_{\text{col}} \| \leq G$, the following holds.
\begin{align}
   \frac{L}{2}\eta_t^2 \|g\|^2 \leq \frac{L}{2}\eta_t^2 G^2
\end{align}

Combined above, we have
\begin{align}
    \|L_{t+1} -L_t \| \leq (2G^2 L + \frac{L}{2}G^2) \eta_t^2 = \frac{5}{2}G^2 L \eta_t^2
\end{align}

Define \(C_0:= \frac{5}{2}\,G^{2}L\) to obtain
\[
|L_{t+1}-L_t|\;\le\;C_0\,\eta_t^{2}.
\]

Because \(\sum_{t=0}^{\infty}\eta_t^{2}<\infty\)
(Robbins–Monro assumption),
\[
\sum_{t\in\mathcal N}|L_{t+1}-L_t|
\;\le\;
C_0\sum_{t\in\mathcal N}\eta_t^{2}
\;\le\;
C_0\sum_{t=0}^{\infty}\eta_t^{2}
\;<\;\infty.
\]

\begin{theorem}
\label{theorem:gap-conv_appendix}
Let $V_{\text{ind}}$ and $V_{\text{col}}$ be $L$-smooth.  
Assume the step size satisfies the Robbins–Monro conditions:  
$0<\eta_t \leq |\delta_t|/L$ with $\sum_t \eta_t = \infty$ and $\sum_t \eta_t^2 < \infty$.  
Also assume conflict recurrence, meaning that for any $\epsilon > 0$ and any $t$, if $|\delta_t| \geq \epsilon$, then there exists $t' \geq t$ such that $(g_{\text{ind}, t'} \cdot g_{\text{col}, t'}) < 0$.  
Then, the value gap converges to zero:
\begin{align}
    \lim_{t \to \infty} |V_{\text{ind}}(\theta_t) - V_{\text{col}}(\theta_t)| = 0.
\end{align}
\end{theorem}

\textit{Proof}. Denote conflict indices by $\mathcal C$ and non‑conflict by $\mathcal N$. Lemma A.3 and Lemma A.4 give for every horizon $T$
\begin{align}
    L_T \;\le\;L_0
-\frac12\!\sum_{t\in\mathcal C,\,t<T}\!\eta_t\,|\delta_t|\,\|d_t\|^{2}
+\;C_0\!\sum_{t\in\mathcal N,\,t<T}\eta_t^{2}.
\end{align}

According to the assumption of the Robbins-Monro, the following holds:
\begin{align}
\sum_{t\in\mathcal C}\eta_t\,|\delta_t|\,\|d_t\|^{2}<\infty.
\end{align}

For any conflict step the projection property and
bounded gradients imply  $\|g_t\|\ge\sigma>0$ with $\sigma:=\tfrac12\min(\|g_{\text{ind},t}\|,\|g_{\text{col},t}\|)$. Thus, we have
\begin{align}\label{eq:contr}
\sum_{t\in\mathcal C}\eta_t\,|\delta_t|\leq \sigma^{-2}\sum_{t\in\mathcal C}\eta_t |\delta_t|\|g_t\|^2 <\infty.
\end{align}

Here, we use contradiction. Assume $\limsup_{t\to\infty}|\delta_t|=\varepsilon_0>0$. Set $\varepsilon:=\varepsilon_0/2$. By the assumption, there exists an \emph{infinite} set
$\mathcal C_\varepsilon
=\{\,t\in\mathcal C\mid |\delta_t|\ge\varepsilon\}.$
Then for every $t\in\mathcal C_\varepsilon$,
$\eta_t\,|\delta_t|\ge\eta_t\,\varepsilon.$ Because $\sum_t \eta_t = \infty$, $\sum_{t\in\mathcal C_\varepsilon}\eta_t\,\varepsilon=\infty$, contradicting the finiteness of Eq. \ref{eq:contr}. Therefore, $\limsup_{t\to\infty}|\delta_t|=0$.

\newpage

\section{Implementation Details}

All experiments were run on a local server equipped with an AMD EPYC 7713 64-Core CPU and five NVIDIA RTX 6000 Ada Generation GPUs. Each rollout consisted of 64–256 parallel environments depending on the task, and training time per run ranged from 2 to 8 hours.

\subsection{Unfair Coin}

Each agent has a CNN-based actor-critic network. The observation is processed through three convolutional layers with kernel sizes of $5 \times 5$, $3 \times 3$, and $3 \times 3$, each with 32 channels and ReLU activations, followed by a fully connected layer with 64 units. The actor head outputs a categorical distribution over discrete actions, while the critic consists of two separate heads estimating the individual and collective value functions.

We train the networks using the Adam optimizer with a learning rate of $1 \times 10^{-4}$, linearly annealed over time. PPO is used with a clipping threshold of 0.2 and two update epochs per iteration, using 500 minibatches. We collect trajectories from 256 parallel environments, each running for 1000 steps per rollout. The discount factor is set to $\gamma = 0.99$ and the GAE parameter to $\lambda = 0.95$. The entropy and value loss coefficients are set to 0.1, respectively. Gradients are clipped to a maximum global norm of 0.5.

\subsection{Cleanup}

Each agent is equipped with a convolutional actor-critical network. The observation is processed through three convolutional layers with kernel sizes of $5 \times 5$, $3 \times 3$, and $3 \times 3$, each with 32 channels and ReLU activations, followed by a fully connected layer with 64 units. The actor outputs a categorical distribution over discrete actions, and the critic consists of two heads that estimate the individual and collective value functions, respectively.

Training is performed using PPO with a clipping threshold of 0.2 and two update epochs per iteration. A total of 500 minibatches are used per update, with data collected from 64 parallel environments running 1000 steps per rollout. The discount factor is set to $\gamma = 0.99$, and the GAE parameter is set to $\lambda = 0.95$. We use the Adam optimizer with an initial learning rate of $5 \times 10^{-4}$, which is linearly annealed during training. The value loss coefficient and entropy coefficient are both set to 0.01, and the value function loss is weighted by 0.5. Gradients are clipped with a maximum global norm of 0.5.

\subsection{Harvest}

Each agent is equipped with a convolutional actor-critical network. The observation is processed through three convolutional layers with kernel sizes of $5 \times 5$, $3 \times 3$, and $3 \times 3$, each with 32 channels and ReLU activations, followed by a fully connected layer with 64 units. The actor outputs a categorical distribution over discrete actions, and the critic consists of two heads that estimate the individual and collective value functions, respectively.

Training is performed using PPO with a clipping threshold of 0.2 and two update epochs per iteration. A total of 500 minibatches are used per update, with data collected from 64 parallel environments running 1000 steps per rollout. The discount factor is set to $\gamma = 0.99$, and the GAE parameter is set to $\lambda = 0.95$. We use the Adam optimizer with an initial learning rate of $5 \times 10^{-4}$, which is linearly annealed during training. The entropy and value function loss coefficients are set to 0.01 and 0.5, respectively. Gradients are clipped to a maximum global norm of 0.5.


\newpage
\section*{NeurIPS Paper Checklist}

\begin{enumerate}

\item {\bf Claims}
    \item[] Question: Do the main claims made in the abstract and introduction accurately reflect the paper's contributions and scope?
    \item[] Answer: \answerYes{} 
    \item[] Justification: The abstract and introduction clearly state the core contributions and are consistent with both theoretical and empirical results.
    \item[] Guidelines:
    \begin{itemize}
        \item The answer NA means that the abstract and introduction do not include the claims made in the paper.
        \item The abstract and/or introduction should clearly state the claims made, including the contributions made in the paper and important assumptions and limitations. A No or NA answer to this question will not be perceived well by the reviewers. 
        \item The claims made should match theoretical and experimental results, and reflect how much the results can be expected to generalize to other settings. 
        \item It is fine to include aspirational goals as motivation as long as it is clear that these goals are not attained by the paper. 
    \end{itemize}

\item {\bf Limitations}
    \item[] Question: Does the paper discuss the limitations of the work performed by the authors?
    \item[] Answer: \answerYes{} 
    \item[] Justification: We stated the limitation in the conclusion.
    \item[] Guidelines:
    \begin{itemize}
        \item The answer NA means that the paper has no limitation while the answer No means that the paper has limitations, but those are not discussed in the paper. 
        \item The authors are encouraged to create a separate "Limitations" section in their paper.
        \item The paper should point out any strong assumptions and how robust the results are to violations of these assumptions (e.g., independence assumptions, noiseless settings, model well-specification, asymptotic approximations only holding locally). The authors should reflect on how these assumptions might be violated in practice and what the implications would be.
        \item The authors should reflect on the scope of the claims made, e.g., if the approach was only tested on a few datasets or with a few runs. In general, empirical results often depend on implicit assumptions, which should be articulated.
        \item The authors should reflect on the factors that influence the performance of the approach. For example, a facial recognition algorithm may perform poorly when image resolution is low or images are taken in low lighting. Or a speech-to-text system might not be used reliably to provide closed captions for online lectures because it fails to handle technical jargon.
        \item The authors should discuss the computational efficiency of the proposed algorithms and how they scale with dataset size.
        \item If applicable, the authors should discuss possible limitations of their approach to address problems of privacy and fairness.
        \item While the authors might fear that complete honesty about limitations might be used by reviewers as grounds for rejection, a worse outcome might be that reviewers discover limitations that aren't acknowledged in the paper. The authors should use their best judgment and recognize that individual actions in favor of transparency play an important role in developing norms that preserve the integrity of the community. Reviewers will be specifically instructed to not penalize honesty concerning limitations.
    \end{itemize}

\item {\bf Theory assumptions and proofs}
    \item[] Question: For each theoretical result, does the paper provide the full set of assumptions and a complete (and correct) proof?
    \item[] Answer: \answerYes{} 
    \item[] Justification: We included the proof in the Appendix.
    \item[] Guidelines:
    \begin{itemize}
        \item The answer NA means that the paper does not include theoretical results. 
        \item All the theorems, formulas, and proofs in the paper should be numbered and cross-referenced.
        \item All assumptions should be clearly stated or referenced in the statement of any theorems.
        \item The proofs can either appear in the main paper or the supplemental material, but if they appear in the supplemental material, the authors are encouraged to provide a short proof sketch to provide intuition. 
        \item Inversely, any informal proof provided in the core of the paper should be complemented by formal proofs provided in appendix or supplemental material.
        \item Theorems and Lemmas that the proof relies upon should be properly referenced. 
    \end{itemize}

    \item {\bf Experimental result reproducibility}
    \item[] Question: Does the paper fully disclose all the information needed to reproduce the main experimental results of the paper to the extent that it affects the main claims and/or conclusions of the paper (regardless of whether the code and data are provided or not)?
    \item[] Answer: \answerYes{} 
    \item[] Justification: We provided the details in the Appendix and the main paper.
    \item[] Guidelines:
    \begin{itemize}
        \item The answer NA means that the paper does not include experiments.
        \item If the paper includes experiments, a No answer to this question will not be perceived well by the reviewers: Making the paper reproducible is important, regardless of whether the code and data are provided or not.
        \item If the contribution is a dataset and/or model, the authors should describe the steps taken to make their results reproducible or verifiable. 
        \item Depending on the contribution, reproducibility can be accomplished in various ways. For example, if the contribution is a novel architecture, describing the architecture fully might suffice, or if the contribution is a specific model and empirical evaluation, it may be necessary to either make it possible for others to replicate the model with the same dataset, or provide access to the model. In general. releasing code and data is often one good way to accomplish this, but reproducibility can also be provided via detailed instructions for how to replicate the results, access to a hosted model (e.g., in the case of a large language model), releasing of a model checkpoint, or other means that are appropriate to the research performed.
        \item While NeurIPS does not require releasing code, the conference does require all submissions to provide some reasonable avenue for reproducibility, which may depend on the nature of the contribution. For example
        \begin{enumerate}
            \item If the contribution is primarily a new algorithm, the paper should make it clear how to reproduce that algorithm.
            \item If the contribution is primarily a new model architecture, the paper should describe the architecture clearly and fully.
            \item If the contribution is a new model (e.g., a large language model), then there should either be a way to access this model for reproducing the results or a way to reproduce the model (e.g., with an open-source dataset or instructions for how to construct the dataset).
            \item We recognize that reproducibility may be tricky in some cases, in which case authors are welcome to describe the particular way they provide for reproducibility. In the case of closed-source models, it may be that access to the model is limited in some way (e.g., to registered users), but it should be possible for other researchers to have some path to reproducing or verifying the results.
        \end{enumerate}
    \end{itemize}

\item {\bf Open access to data and code}
    \item[] Question: Does the paper provide open access to the data and code, with sufficient instructions to faithfully reproduce the main experimental results, as described in supplemental material?
    \item[] Answer: \answerYes{} 
    \item[] Justification: We cited the corresponding paper.
    \item[] Guidelines:
    \begin{itemize}
        \item The answer NA means that paper does not include experiments requiring code.
        \item Please see the NeurIPS code and data submission guidelines (\url{https://nips.cc/public/guides/CodeSubmissionPolicy}) for more details.
        \item While we encourage the release of code and data, we understand that this might not be possible, so “No” is an acceptable answer. Papers cannot be rejected simply for not including code, unless this is central to the contribution (e.g., for a new open-source benchmark).
        \item The instructions should contain the exact command and environment needed to run to reproduce the results. See the NeurIPS code and data submission guidelines (\url{https://nips.cc/public/guides/CodeSubmissionPolicy}) for more details.
        \item The authors should provide instructions on data access and preparation, including how to access the raw data, preprocessed data, intermediate data, and generated data, etc.
        \item The authors should provide scripts to reproduce all experimental results for the new proposed method and baselines. If only a subset of experiments are reproducible, they should state which ones are omitted from the script and why.
        \item At submission time, to preserve anonymity, the authors should release anonymized versions (if applicable).
        \item Providing as much information as possible in supplemental material (appended to the paper) is recommended, but including URLs to data and code is permitted.
    \end{itemize}

\item {\bf Experimental setting/details}
    \item[] Question: Does the paper specify all the training and test details (e.g., data splits, hyperparameters, how they were chosen, type of optimizer, etc.) necessary to understand the results?
    \item[] Answer: \answerYes{} 
    \item[] Justification: We provided the implementation details in the Appendix.
    \item[] Guidelines:
    \begin{itemize}
        \item The answer NA means that the paper does not include experiments.
        \item The experimental setting should be presented in the core of the paper to a level of detail that is necessary to appreciate the results and make sense of them.
        \item The full details can be provided either with the code, in appendix, or as supplemental material.
    \end{itemize}

\item {\bf Experiment statistical significance}
    \item[] Question: Does the paper report error bars suitably and correctly defined or other appropriate information about the statistical significance of the experiments?
    \item[] Answer: \answerYes{} 
    \item[] Justification: We provided the mean and variance of individual returns.
    \item[] Guidelines:
    \begin{itemize}
        \item The answer NA means that the paper does not include experiments.
        \item The authors should answer "Yes" if the results are accompanied by error bars, confidence intervals, or statistical significance tests, at least for the experiments that support the main claims of the paper.
        \item The factors of variability that the error bars are capturing should be clearly stated (for example, train/test split, initialization, random drawing of some parameter, or overall run with given experimental conditions).
        \item The method for calculating the error bars should be explained (closed form formula, call to a library function, bootstrap, etc.)
        \item The assumptions made should be given (e.g., Normally distributed errors).
        \item It should be clear whether the error bar is the standard deviation or the standard error of the mean.
        \item It is OK to report 1-sigma error bars, but one should state it. The authors should preferably report a 2-sigma error bar than state that they have a 96\% CI, if the hypothesis of Normality of errors is not verified.
        \item For asymmetric distributions, the authors should be careful not to show in tables or figures symmetric error bars that would yield results that are out of range (e.g. negative error rates).
        \item If error bars are reported in tables or plots, The authors should explain in the text how they were calculated and reference the corresponding figures or tables in the text.
    \end{itemize}

\item {\bf Experiments compute resources}
    \item[] Question: For each experiment, does the paper provide sufficient information on the computer resources (type of compute workers, memory, time of execution) needed to reproduce the experiments?
    \item[] Answer: \answerYes{} 
    \item[] Justification: We stated this in the Appendix.
    \item[] Guidelines:
    \begin{itemize}
        \item The answer NA means that the paper does not include experiments.
        \item The paper should indicate the type of compute workers CPU or GPU, internal cluster, or cloud provider, including relevant memory and storage.
        \item The paper should provide the amount of compute required for each of the individual experimental runs as well as estimate the total compute. 
        \item The paper should disclose whether the full research project required more compute than the experiments reported in the paper (e.g., preliminary or failed experiments that didn't make it into the paper). 
    \end{itemize}
    
\item {\bf Code of ethics}
    \item[] Question: Does the research conducted in the paper conform, in every respect, with the NeurIPS Code of Ethics \url{https://neurips.cc/public/EthicsGuidelines}?
    \item[] Answer: \answerYes{} 
    \item[] Justification: We follow the NeurIPS Code of Ethics.
    \item[] Guidelines:
    \begin{itemize}
        \item The answer NA means that the authors have not reviewed the NeurIPS Code of Ethics.
        \item If the authors answer No, they should explain the special circumstances that require a deviation from the Code of Ethics.
        \item The authors should make sure to preserve anonymity (e.g., if there is a special consideration due to laws or regulations in their jurisdiction).
    \end{itemize}

\item {\bf Broader impacts}
    \item[] Question: Does the paper discuss both potential positive societal impacts and negative societal impacts of the work performed?
    \item[] Answer: \answerYes{} 
    \item[] Justification: We stated the broader impacts in the conclusion.
    \item[] Guidelines:
    \begin{itemize}
        \item The answer NA means that there is no societal impact of the work performed.
        \item If the authors answer NA or No, they should explain why their work has no societal impact or why the paper does not address societal impact.
        \item Examples of negative societal impacts include potential malicious or unintended uses (e.g., disinformation, generating fake profiles, surveillance), fairness considerations (e.g., deployment of technologies that could make decisions that unfairly impact specific groups), privacy considerations, and security considerations.
        \item The conference expects that many papers will be foundational research and not tied to particular applications, let alone deployments. However, if there is a direct path to any negative applications, the authors should point it out. For example, it is legitimate to point out that an improvement in the quality of generative models could be used to generate deepfakes for disinformation. On the other hand, it is not needed to point out that a generic algorithm for optimizing neural networks could enable people to train models that generate Deepfakes faster.
        \item The authors should consider possible harms that could arise when the technology is being used as intended and functioning correctly, harms that could arise when the technology is being used as intended but gives incorrect results, and harms following from (intentional or unintentional) misuse of the technology.
        \item If there are negative societal impacts, the authors could also discuss possible mitigation strategies (e.g., gated release of models, providing defenses in addition to attacks, mechanisms for monitoring misuse, mechanisms to monitor how a system learns from feedback over time, improving the efficiency and accessibility of ML).
    \end{itemize}
    
\item {\bf Safeguards}
    \item[] Question: Does the paper describe safeguards that have been put in place for responsible release of data or models that have a high risk for misuse (e.g., pretrained language models, image generators, or scraped datasets)?
    \item[] Answer: \answerNA{} 
    \item[] Justification: \answerNA{}
    \item[] Guidelines:
    \begin{itemize}
        \item The answer NA means that the paper poses no such risks.
        \item Released models that have a high risk for misuse or dual-use should be released with necessary safeguards to allow for controlled use of the model, for example by requiring that users adhere to usage guidelines or restrictions to access the model or implementing safety filters. 
        \item Datasets that have been scraped from the Internet could pose safety risks. The authors should describe how they avoided releasing unsafe images.
        \item We recognize that providing effective safeguards is challenging, and many papers do not require this, but we encourage authors to take this into account and make a best faith effort.
    \end{itemize}

\item {\bf Licenses for existing assets}
    \item[] Question: Are the creators or original owners of assets (e.g., code, data, models), used in the paper, properly credited and are the license and terms of use explicitly mentioned and properly respected?
    \item[] Answer: \answerYes{} 
    \item[] Justification: We have cited the paper.
    \item[] Guidelines:
    \begin{itemize}
        \item The answer NA means that the paper does not use existing assets.
        \item The authors should cite the original paper that produced the code package or dataset.
        \item The authors should state which version of the asset is used and, if possible, include a URL.
        \item The name of the license (e.g., CC-BY 4.0) should be included for each asset.
        \item For scraped data from a particular source (e.g., website), the copyright and terms of service of that source should be provided.
        \item If assets are released, the license, copyright information, and terms of use in the package should be provided. For popular datasets, \url{paperswithcode.com/datasets} has curated licenses for some datasets. Their licensing guide can help determine the license of a dataset.
        \item For existing datasets that are re-packaged, both the original license and the license of the derived asset (if it has changed) should be provided.
        \item If this information is not available online, the authors are encouraged to reach out to the asset's creators.
    \end{itemize}

\item {\bf New assets}
    \item[] Question: Are new assets introduced in the paper well documented and is the documentation provided alongside the assets?
    \item[] Answer: \answerNA{} 
    \item[] Justification: \answerNA{}
    \item[] Guidelines:
    \begin{itemize}
        \item The answer NA means that the paper does not release new assets.
        \item Researchers should communicate the details of the dataset/code/model as part of their submissions via structured templates. This includes details about training, license, limitations, etc. 
        \item The paper should discuss whether and how consent was obtained from people whose asset is used.
        \item At submission time, remember to anonymize your assets (if applicable). You can either create an anonymized URL or include an anonymized zip file.
    \end{itemize}

\item {\bf Crowdsourcing and research with human subjects}
    \item[] Question: For crowdsourcing experiments and research with human subjects, does the paper include the full text of instructions given to participants and screenshots, if applicable, as well as details about compensation (if any)? 
    \item[] Answer: \answerNA{} 
    \item[] Justification: \answerNA{}
    \item[] Guidelines:
    \begin{itemize}
        \item The answer NA means that the paper does not involve crowdsourcing nor research with human subjects.
        \item Including this information in the supplemental material is fine, but if the main contribution of the paper involves human subjects, then as much detail as possible should be included in the main paper. 
        \item According to the NeurIPS Code of Ethics, workers involved in data collection, curation, or other labor should be paid at least the minimum wage in the country of the data collector. 
    \end{itemize}

\item {\bf Institutional review board (IRB) approvals or equivalent for research with human subjects}
    \item[] Question: Does the paper describe potential risks incurred by study participants, whether such risks were disclosed to the subjects, and whether Institutional Review Board (IRB) approvals (or an equivalent approval/review based on the requirements of your country or institution) were obtained?
    \item[] Answer: \answerNA{} 
    \item[] Justification: \answerNA{}
    \item[] Guidelines:
    \begin{itemize}
        \item The answer NA means that the paper does not involve crowdsourcing nor research with human subjects.
        \item Depending on the country in which research is conducted, IRB approval (or equivalent) may be required for any human subjects research. If you obtained IRB approval, you should clearly state this in the paper. 
        \item We recognize that the procedures for this may vary significantly between institutions and locations, and we expect authors to adhere to the NeurIPS Code of Ethics and the guidelines for their institution. 
        \item For initial submissions, do not include any information that would break anonymity (if applicable), such as the institution conducting the review.
    \end{itemize}

\item {\bf Declaration of LLM usage}
    \item[] Question: Does the paper describe the usage of LLMs if it is an important, original, or non-standard component of the core methods in this research? Note that if the LLM is used only for writing, editing, or formatting purposes and does not impact the core methodology, scientific rigorousness, or originality of the research, declaration is not required.
    \item[] Answer: \answerNA{} 
    \item[] Justification: \answerNA{}
    \item[] Guidelines:
    \begin{itemize}
        \item The answer NA means that the core method development in this research does not involve LLMs as any important, original, or non-standard components.
        \item Please refer to our LLM policy (\url{https://neurips.cc/Conferences/2025/LLM}) for what should or should not be described.
    \end{itemize}

\end{enumerate}

\end{document}